\documentclass[prl,twocolumn,amsfonts,amssymb,amsmath,floatfix,tightenlines,superscriptaddress]{revtex4-1}
\usepackage{amsfonts,amssymb}
\usepackage[subnum]{cases}
\usepackage{mathrsfs}
\usepackage{amsmath}
\usepackage[none]{hyphenat}
\usepackage{graphicx}
\usepackage{titlesec}
\usepackage{hyperref}    % links
\usepackage{footmisc}%\usepackage{cleveref}
\usepackage{bm}          % bold math letters
\usepackage{natbib}
\usepackage{soul} % in order to highlight
\usepackage{color}
\usepackage{longtable}
\usepackage{ textcomp }
\usepackage{verbatim}
\usepackage{lipsum}
\usepackage{titlecaps}

\titleformat{\section}
    [block]{\bfseries\normalsize}{\rlap{\thesection}}{\titlecap}
    {\hspace{.0\linewidth}\begin{minipage}[t]{.85\textwidth}}[\end{minipage}]
\titleformat{\subsection}
    [block]{\bfseries\normalsize}{\rlap{\thesection}}{\titlecap}
    {\hspace{.0\linewidth}\begin{minipage}[t]{.85\textwidth}}[\end{minipage}]

\newcommand{\beginsupplement}{%
        \setcounter{table}{0}
        \renewcommand{\thetable}{S\arabic{table}}%
        \setcounter{figure}{0}
        \renewcommand{\thefigure}{S\arabic{figure}}%
        \renewcommand{\thesection}{\Roman{section}} %
        \setcounter{equation}{0}
        \renewcommand{\theequation}{S\arabic{equation}}%
     }

\begin{document}

\title{Jet Sub-structure in  Fireworks Emission from \\ Non-uniform and Rotating Bose-Einstein Condensates}
\author{Han Fu}
\email{vickeyrobert@uchicago.edu}
\affiliation{James Franck Institute, University of Chicago, Chicago, IL 60637, USA}
\affiliation{Department of Physics, University of Chicago, Chicago, IL 60637, USA}
\author{Zhendong Zhang}
\affiliation{James Franck Institute, University of Chicago, Chicago, IL 60637, USA}
\affiliation{Department of Physics, University of Chicago, Chicago, IL 60637, USA}
\affiliation{Enrico Fermi Institute and Department of Physics, University of Chicago, Chicago, IL 60637, USA}
\author{Kai-Xuan Yao}
\affiliation{James Franck Institute, University of Chicago, Chicago, IL 60637, USA}
\affiliation{Department of Physics, University of Chicago, Chicago, IL 60637, USA}
\affiliation{Enrico Fermi Institute and Department of Physics, University of Chicago, Chicago, IL 60637, USA}
\author{Lei Feng}
\affiliation{James Franck Institute, University of Chicago, Chicago, IL 60637, USA}
\affiliation{Department of Physics, University of Chicago, Chicago, IL 60637, USA}
\affiliation{Enrico Fermi Institute and Department of Physics, University of Chicago, Chicago, IL 60637, USA}
\author{Jooheon Yoo}
\affiliation{Department of Physics, University of Chicago, Chicago, IL 60637, USA}
\author{Logan W. Clark}
\affiliation{James Franck Institute, University of Chicago, Chicago, IL 60637, USA}
\affiliation{Department of Physics, University of Chicago, Chicago, IL 60637, USA}
\affiliation{Enrico Fermi Institute and Department of Physics, University of Chicago, Chicago, IL 60637, USA}
\author{K. Levin}
\affiliation{James Franck Institute, University of Chicago, Chicago, IL 60637, USA}
\affiliation{Department of Physics, University of Chicago, Chicago, IL 60637, USA}
\author{Cheng Chin}
\affiliation{James Franck Institute, University of Chicago, Chicago, IL 60637, USA}
\affiliation{Department of Physics, University of Chicago, Chicago, IL 60637, USA}
\affiliation{Enrico Fermi Institute and Department of Physics, University of Chicago, Chicago, IL 60637, USA}
\date{\today}

\begin{abstract}
We show that jet emission from a Bose condensate with
periodically driven interactions, a.k.a. ``Bose fireworks", contains essential
information on the condensate wavefunction, which is difficult to obtain using standard detection methods. We illustrate the underlying physics with two examples. When condensates acquire phase patterns from external potentials or from vortices, the jets display novel sub-structure, such as oscillations or spirals, in their correlations. Through a comparison of theory, numerical simulations and experiments, we show how one can quantitatively extract the phase and the helicity of a condensate from the emission pattern. Our work demonstrating the strong link between jet emission and the underlying quantum system, bears
on the recent emphasis on jet sub-structure in particle physics.
\end{abstract}

\maketitle
Cold atom systems are emerging as an important platform for
quantum simulations
in condensed matter \cite{KLevin} and in high
energy physics \cite{Arratia}.
Of current interest is Floquet engineering, the application of temporal periodic drive to a system.
This has been employed to discover novel phenomena \cite{Eckardt,Review_Cooper} including topological
phases \cite{GoldmanNature,Esslinger} and dynamical gauge fields for simulation of high energy physics models \cite{Logan_gauge,Bloch,gorg2019realization}. \begin{comment}It has attracted much attention in ultracold atomic gases \cite{Eckardt,Review_Cooper},
as it provides promising routes to explore new physical phenomena, including the creation of topological
phases \cite{GoldmanNature,Esslinger} and dynamical gauge fields for simulation of high energy physics models \cite{Logan_gauge,Bloch,gorg2019realization}.
\end{comment}
With driven Bose-Einstein condensates (BECs), a new regime of quantum scattering has been recently reported \cite{Cheng_2017,Slovenia}. Here, periodic control of the atomic interactions excites pairs of atoms propagating in opposite directions. When the modulation strength exceeds a threshold, thin jets of atoms are expelled from the condensate in all directions through Bose stimulation (Bose fireworks). Rich physics can be found in this process such as complex correlations \cite{Lei, Zhai, Yan,Holland}, simulation of Unruh radiation \cite{Unruh}, and density wave formation \cite{DensityWave,zhang2019pattern}.

\begin{figure}[h]
\includegraphics[width=0.5\textwidth]
{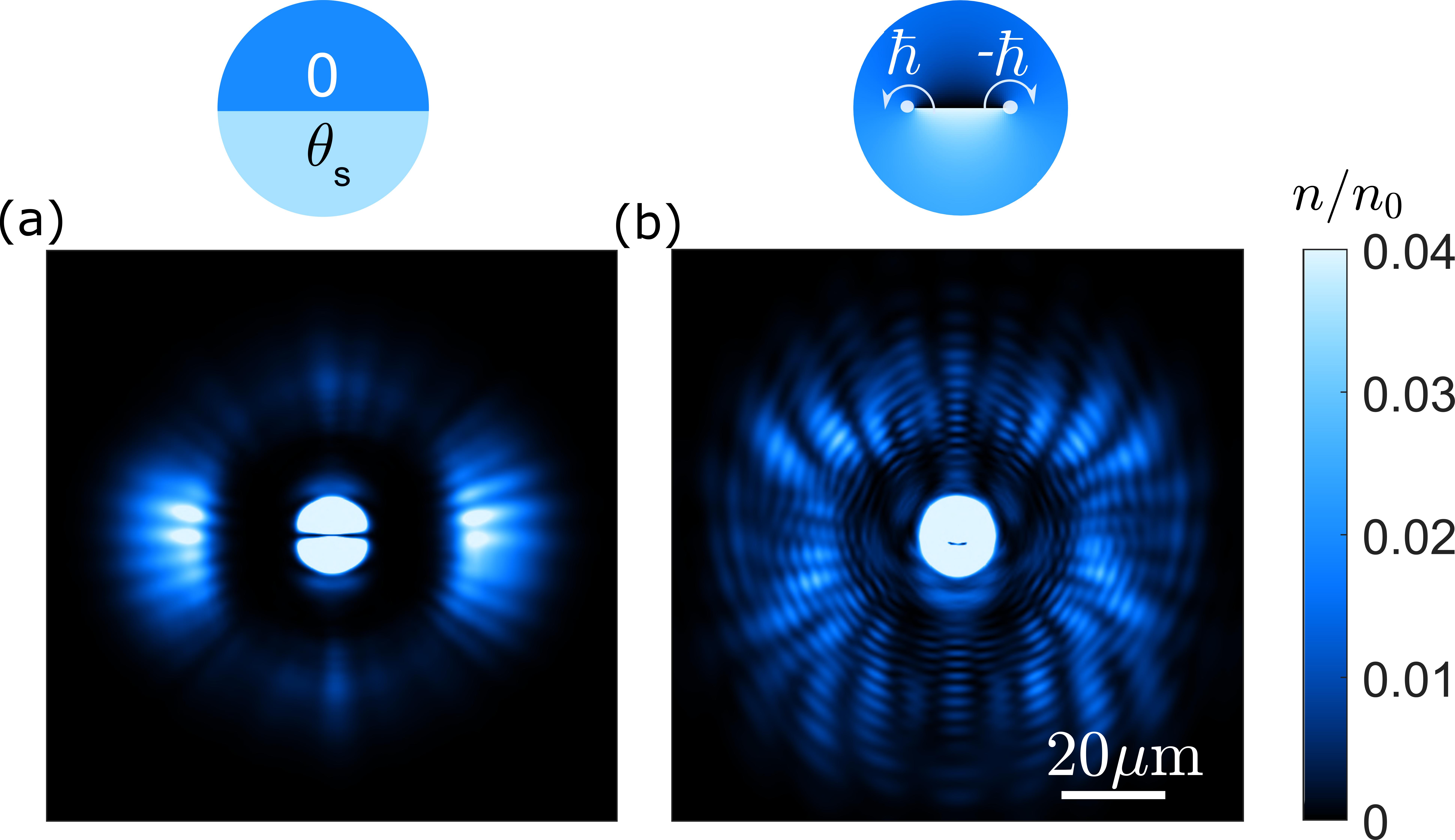}
\caption{Simulated emission patterns of BECs with interaction modulations. (a) Jet emission from condensates with a soliton. The lower half is phase shifted relative to the upper half by $\theta_s=\pi$. (b) Jet emission from condensates with a vortex-antivortex pair, where the reduced Planck constant $\hbar $
corresponds to the angular momentum of the vortex. The atomic density $n$ is normalized to the initial density of the condensate $n_0$.}
\label{fig:soliton}
\end{figure}

In this paper we show how this jet emission pattern of Bose fireworks can enable extraction of the condensate wavefunction. Such studies of jet substructure
are
reminiscent of current
scattering experiments in particle physics performed at
both the Large Hadron \cite{RMP91_045003}  and the Relativistic
Heavy Ion \cite{RMP90_025005} Colliders. It should be pointed out in this regard,
that vorticity (a topic of interest here) is an active sub-field in particle physics
\cite{Natureswirls}. Quark-gluon plasmas exhibiting anomalously high vorticity have
been reported based on
the structure of the particle emission.
To illustrate this capability with cold atoms, a set of emission patterns from numerical simulations are shown in Fig.~\ref{fig:soliton}, which exhibit distinct structures for condensates with different non-uniform phase configurations.

We present two cases of study both experimentally and numerically. In the first, we consider condensates split into two halves with different phases.
The relative phase emerges in the correlations of counter-propagating jets, and can be understood based on the double-slit interference of matterwaves. In the second case, we study condensates with vortices. Here the emission
pattern exhibits a novel spiral sub-structure as seen in
Fig.~\ref{fig:soliton}(b). We show that one can directly extract the phase winding number of the vortices from the spirals. Excellent agreement between experiments and simulations is obtained for both cases.

Our study suggests jet emission as a new tool to probe the condensate phase distribution, which can be difficult to access with conventional detection schemes such as in situ and time-of-flight imaging. These often involve multiple stages of experimental preparation.
For instance, a complex setup is required to visualize vortices \cite{Cornell_vortex,Ketterle_TOF,Anderson2015} or to measure the helicity of vortices \cite{Vor_img} such as by interfering two condensates \cite{Dalibard2,Ketterle_vortex}. Our method can reveal phase information such as the helicity of a
vortex, and, moreover, does not necessitate destroying
the entire condensate while imaging. In principle,
subsequent imaging processes offer the possibility to follow more detailed
evolutionary dynamics.
These techniques can also be generalized to atoms in optical lattices.

In our simulations, we describe the evolution of the condensates with the Gross-Pitaevskii (GP) equation, including terms that simulate quantum fluctuations \cite{DensityWave}.
For a uniform BEC, periodic modulation of the interaction strength with frequency $\omega$ leads to pair production of matterwave jets with random but opposite momenta $(\hbar{\bf{k_f}},-\hbar\bf{k_f})$, where $k_f=\sqrt{m\omega/\hbar}$ and $m$ is the atomic mass. For non-uniform condensates, jets form in pairs of modes which are determined by the condensate wavefunction and driving frequency. When observed in the plane wave basis, the jets can show intricate correlations. The goal of this work is to demonstrate that much can be learned about the condensate from the strength and correlations of the emitted jets.

Microscopically, the system under the periodic drive is excited from an initial state $\psi_0$ to  $\psi(t)\equiv\psi_0+\delta\psi$, where the wavefunction increment $\delta\psi$ can be seeded by quantum fluctuations and amplified by the drive. With short interaction times as in our experiment, the deviation can be treated perturbatively, and the evolution of the system is governed by the Hamiltonian
\begin{equation}\label{eq:ham}
\begin{aligned}
H \approx&\sum_i  E_i a_{i}^\dagger
a_{i}+\frac{U(t)}{2}\sum_{i,i^\prime}\left[F(i,i^\prime)a_{i}^\dagger a_{i^\prime}^\dagger+ h.c.\right],\\
\end{aligned}
\end{equation}
where $U(t)=U_0+U_1\sin\omega t$ is the oscillating interaction strength, $\sum_i$ sums over the single particle modes $\varphi_i$ that are initially unoccupied,  the pair function $F(i,i')$ is described below, $E_i$ is the kinetic energy of the $i-$th mode, and
$a_i$ and $a_i^\dagger$ are the annihilation and creation operators of the mode. Here we work in the regime where the modulation amplitude is much larger than the offset, and the driving energy is much greater than the energy of the initial state, i.e. $U_0 n_0\ll U_1n_0\ll\hbar\omega$, where $n_0$ is the average density of the condensate \footnote{In the high frequency limit, additional terms like $U(t)\int dx dy |\psi_0(x,y)|^2\varphi_i^*(x,y)\varphi_{i^\prime}(x,y)$ is negligible, see Ref. \onlinecite{Logan}.}.

The pair function $F(i,i^\prime)$ in Eq.~(1) determines the strength as
well as the correlations of the two modes $i$ and $i^\prime$ in the emission. It is given by the overlap of the condensate wavefunction $\psi_0$ and the wavefunctions of the modes
$\varphi_i$ and $\varphi_{i^\prime}$, namely,
\begin{equation}\label{eq:funcF}
\begin{aligned}
F(i,i^\prime) &=\int
d\textbf{r} \varphi_i^* (\textbf{r})\varphi_{i^\prime}^*(\textbf{r})\psi_0^2(\textbf{r}).
\end{aligned}
\end{equation}

This equation shows that, in principle we can determine the square of the condensate wavefunction directly from the pair
function $F$.
As an example,
if we choose a plane wave basis, $F({\bf{k}},{\bf{k^\prime}})$ is the ${\bf{k}}+{\bf{k^\prime}}$ Fourier component of $\psi_0^2$. When the condensate contains multiple excitations, those with larger amplitudes of $F({\bf{k}},{\bf{k^\prime}})$ will lead to stronger emission of the matterwave jets with momenta $\bf{k}$ and $\bf{k^\prime}$, providing
they satisfy momentum conservation conditions.
The precise mathematics and procedure to extract the pair function $F$ from correlations in firework emissions is provided in the Supplement Sec. 4 \cite{Supple}. To validate these
ideas and offer a physical picture we study two examples of non-uniform BECs experimentally and theoretically. These cases involve BECs with two different phases and with vortices; both of which illustrate the links between the jet substructure and condensate wavefunctions through the comparisons between experiments and simulations.

\begin{figure*}
\includegraphics[width=0.9\textwidth]
{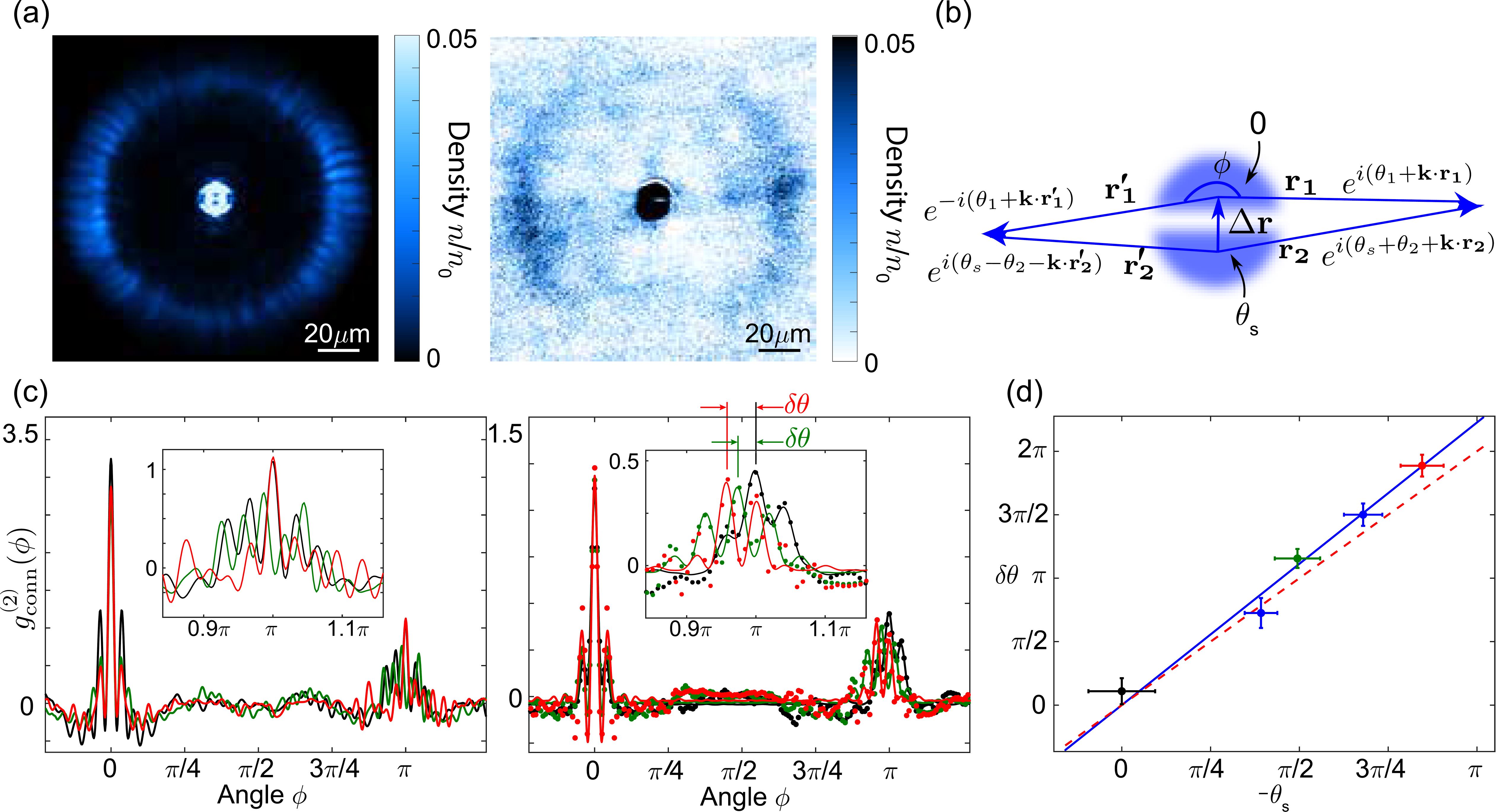}
\caption{Emissions from Bose condensates split into two halves with
and without a relative phase $\theta_s$. (a) Emission pattern for $\theta_s=0$ from  simulations (left) and experiments (right). The density is normalized by the initial average density $n_0$.
(b) Physical picture for the jet emission from two halves of the condensate. (c) Connected correlations $g^{(2)}_{\mathrm{conn}} (\phi)$ for different relative angle $\phi$ (left: simulation, right: experiment) for $\theta_s = 0$ (black), $-\pi/2$ (green), and $-\pi$ (red). Solid lines are fits using the product of a sinc envelope and a sinusoidal function, borrowed from the double-slit interference model. See Supplement Sec. 2 \cite{Supple}. The insets show the jet-substructure of the $\pi$-peaks. The phase shift $\delta\theta$ of the oscillations are indicated by the arrows.
(d) Phase associated with $\pi$-peak shift $\delta\theta$ as a function of imprinted phase $\theta_s$ (plotted against $-\theta_s$). Dots with error bars are experimental data. The blue solid line is a linear fit without intercept and the red dashed line is the theory expectation $\delta\theta=-2\theta_s$, which is identical to the simulation results. Here, the data used in panel (c) are marked out with the same  corresponding colors. Error bars represent 1-$\sigma$ standard deviation.}
\label{fig:artrap}
\end{figure*}

A split BEC with two phases is our first, pedagogical example. A soliton-like structure arises where the phase jump occurs, and the condensate density is suppressed at the boundary. The advantage of considering a
split BEC is that
we are able to disentangle density and phase information which
are strongly intertwined in the soliton case. We assume that at time $t=0$ the phase of the lower half is $\theta_s$ and that of the upper half is zero, and the phase slip boundary is along the $x$-axis. \begin{comment} The wavefunction is then given by
\begin{equation}
\label{eq:dark}
\begin{aligned}
\psi_0 (x,y)=&\sqrt{n_0}e^{i\theta_s/2}
\left\{\cos\frac{\theta_s}{2}\right.\\
&\left.-i\sin\frac{\theta_s}{2}\tanh\left[\sin\frac{\theta_s}{2}\left(\frac{y}{\sqrt{2}\xi}-\frac{\mu t}{\hbar}\cos\frac{\theta_s}{2}\right)\right]\right\}\\
\end{aligned}
\end{equation}
where the phase of the upper half is set as zero and $\xi$ is the healing length set
by the chemical potential $\mu$ as $\hbar^2/2m\xi^2=\mu$.
\end{comment} When the phase shift $\theta_s$ is tuned to $\pi$, the soliton is stationary and the density at the phase boundary reaches zero. Numerical simulation with $\theta_s=\pi$, see Fig.~\ref{fig:soliton}(a), indicates a notable directionality in the emission along the boundary. By comparing the simulation and experiments on split BECs with different phases, we find that the preferred direction of emission comes predominately from the density depletion along the phase boundary and is insensitive to the phases. Jet emission is weakened along the paths that cross the boundary because matterwave amplification ceases with low atomic density.

To experimentally prepare this condensate with two phases, we start with a BEC of $4\times10^4$ cesium atoms in a circular box trap with diameter 18~$\mathrm{\mu m}$ \cite{Cheng_2017}. The sample is tightly confined in the vertical direction with $1/e^2$ radius 0.8 $\mathrm{\mu m}$. We then slowly raise a 6-$\mathrm{\mu m}$ wide potential barrier with a barrier height of $h\times52$~Hz,
thus maintaining phase coherence while substantially separating the BEC into two halves. A phase difference between the two halves is introduced by applying a short light pulse of duration $\tau = 0.4$~ms on one of them. The imprinted phase of $\theta_s = -V_s\tau/\hbar$, where $V_s$ is the light shift, is controlled by the intensity of the light pulse. We calibrate the imprinted phase by interfering the two halves of the BEC after free expansion \cite{Supple}. In the experiment, the potential barrier and the relative phase are controlled independently.

After phase imprinting, we apply an oscillating magnetic field in the vicinity of a Feshbach resonance to initiate the jet emission \cite{Cheng_2017}. The magnetic field modulates the atomic $s$-wave scattering length as $a(t) = a_{dc} + a_{ac}\sin\omega t)$ at frequency $\omega=2\pi\times2.1 $~kHz with a small offset $a_{dc} = 9$~$a_0$ and a large amplitude $a_{ac} = 47$~$a_0$, where $a_0$ is the Bohr radius. The resulting chemical potential is around $h \times 89$ Hz. After the modulation, we perform imaging to record the jets. Emission patterns from experiments and from simulations based on identical parameters show good agreement, see  Fig.~\ref{fig:artrap}(a). This figure illustrates
the fact that the anisotropy in the emission pattern is caused by the density depletion. To see the relative phase one needs to
address the correlations.

We show below how this phase information can be quantitatively extracted.
The phase difference between the two halves $\theta_s$ is revealed in the correlation between counter-propagating jets. We first calculate the connected correlation function $g^{(2)}_{conn}$, defined as
\begin{equation}
g^{(2)}_{conn}(\phi)=\frac{\langle\int_0^\pi d\phi_1 \Delta n_{\phi_1}\Delta n_{\phi_1+\phi}\rangle}{\pi\bar{n}^2},
\end{equation}
where $\Delta x=x-\langle x\rangle$ represents the fluctuation around the mean value, $n_{\phi}$ is the density of the emitted atoms at angle $\phi$, $\langle \cdot\rangle$ denotes the average over all images and $\bar{n}$ is the average density over all directions and images. The correlation function displays a strong peak at $\phi\approx\pi$, called the $\pi$-peak, which indicates that jets form in pairs in opposite propagating directions.

Close examination shows that the $\pi$-peak contains fine oscillations (jet sub-structure) that depend on the condensate phase, see Fig.~\ref{fig:artrap}(c). The phase of the oscillations is found to be proportional to the relative phase between the two halves $\theta_s$. Comparing the phase $\delta \theta$ of the fine oscillations to the phase difference $\theta_s$, we find a linear dependence with a slope -2.2(2), see Fig.~\ref{fig:artrap}(d).
Although there is an uncertainty in the experiments which reflects
calibration errors in the imprinted phase, these measurements
are consistent with the theoretical prediction:

\begin{equation}
\delta\theta=-2\theta_s.
\end{equation}

We provide an intuitive picture to understand this phase relation. In the far field, emission from the upper BEC with probability amplitude $e^{i\left(\theta_1+{\bf{k}}\cdot{\bf{r_1}}\right)}$ propagating to the right overlaps with the emission from the lower half with amplitude $e^{i\left(\theta_s+\theta_2+{\bf{k}}\cdot{\bf{r_2}}\right)}$, where $\theta_1$ and $\theta_2$ are random phases determined by quantum fluctuations, $\bf{k}$ is the jet wavevector and  ${\bf{r_1}}$ ($\bf{r_2}$) is the displacement vector toward the measurement point, see Fig.~\ref{fig:artrap}(b). The two matterwaves interfere and produce a density wave of $\cos(\Delta \theta-\theta_s+{\bf{k}}\cdot\Delta{\bf{ r}})$, where $\Delta \theta=\theta_1-\theta_2$ and $\Delta {\bf{r}}={\bf{r_1}}-{\bf{r_2}}$. Similarly, the left-propagating emissions of amplitudes $e^{i\left(-\theta_1-{\bf{k}}\cdot{\bf{r_1^\prime}}\right)}$ and $e^{i\left(\theta_s-\theta_2-{\bf{k}}\cdot{\bf{r_2^\prime}}\right)}$ overlap and result in a density wave $\cos(\Delta \theta+\theta_s+{\bf{k}}\cdot\Delta{\bf{r}})$. \footnote{From Eq.~(2) the phases of the jets emitted from the same half of the BEC sum to twice the phase of the BEC.}. Comparing the two density waves, we see that the counter-propagating emissions are correlated with a relative phase shift of $\delta\theta=-2\theta_s$.

\begin{figure}[h]
\includegraphics[width=0.48\textwidth]
{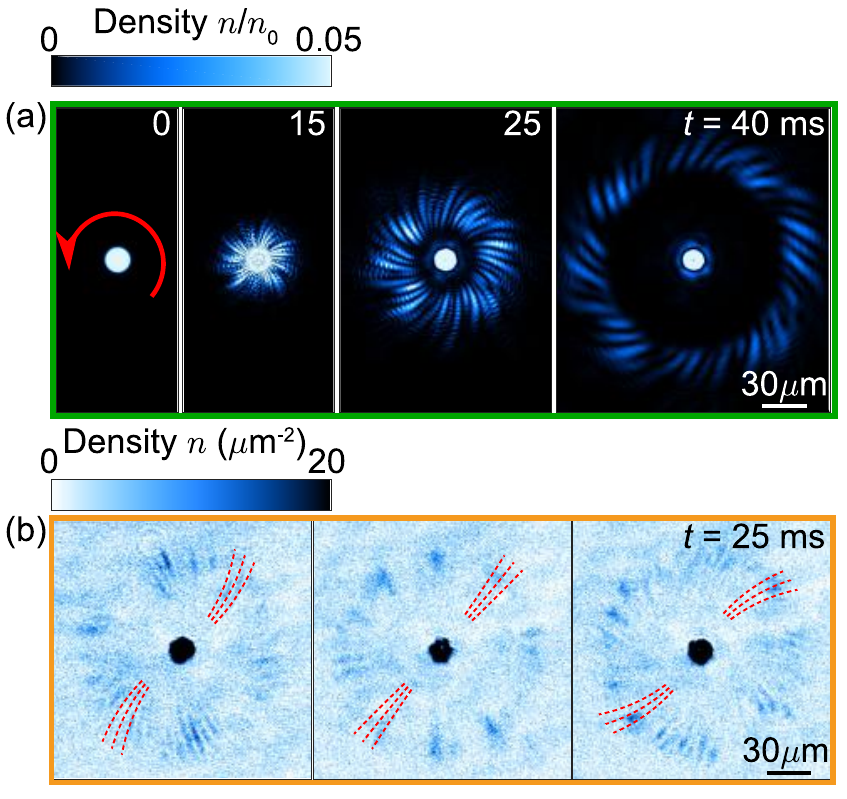}
\caption{Spiral emissions from vortex-embedded BECs. (a) Evolution of the fireworks emission for $\omega/2\pi=2$ kHz and $l_0=1$ from GP simulation. The red arrow indicates the direction of the phase winding with $l_0=1$. (b) Experimental images for $\omega/2
\pi=3$ kHz at $t=25$ ms from BECs with different vortex winding numbers $l_0 = -1,0,1$ from left to right. The red dashed lines are guides to the eye, the curvature of which is calculated from the correlation function \cite{Supple}.}
\label{fig:vortexr}
\end{figure}

The second case study involves vortex-embedded BECs, where the resulting emission patterns display exotic spirals.
In our system the initial condensate wavefunction is characterized  by
an integer winding number $l_0=  \pm1,\,\pm2,\ldots$ as
\begin{equation}\label{eq:vortex} \psi_0(r,\phi)=
\sqrt{n_0(r)}e^{il_0\phi} \end{equation}
in polar coordinates $(r,\phi)$. Since the healing length $\xi$ (set
by the chemical potential $\mu$ as $\hbar^2/2m\xi^2=\mu$) is much smaller than the trap radius $R$, the condensate wavefunction is uniform outside the vortex core. Jet emission dynamics from a driven BEC with a vortex is simulated in Fig.~\ref{fig:vortexr} (a).

In our experiment, about $5\%$ of  condensates form with a vortex. When the system reaches equilibrium, the vortex is expected to settle at the trap center. BECs with and without a vortex can be distinguished from the emission pattern, see Fig.~\ref{fig:vortexr}(b) for emission from BECs with different vorticity \cite{Supple}.

\begin{figure}
\includegraphics[width=.5\textwidth]
{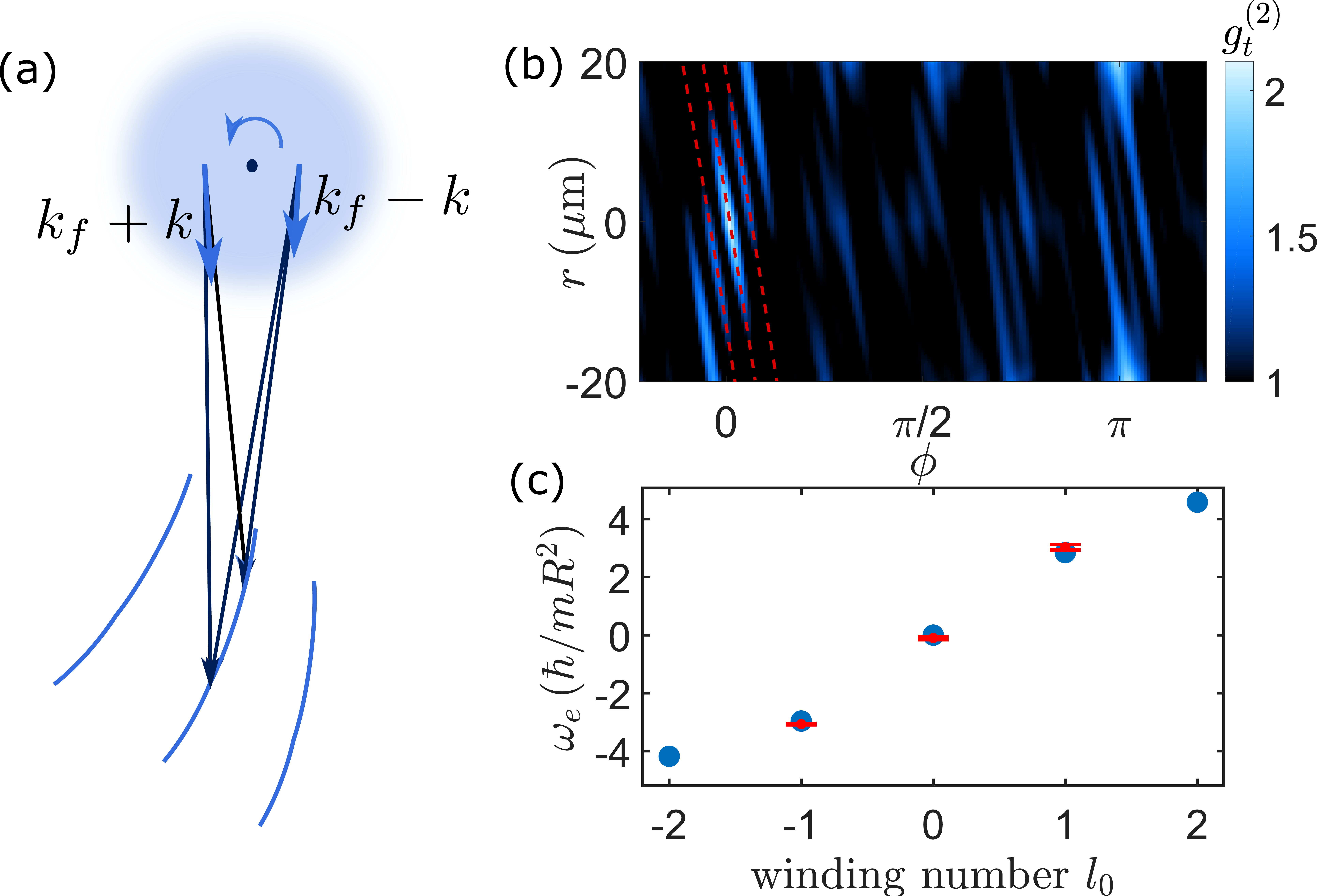}
\caption{Correlation analysis of spiral emission patterns.
(a) A physical picture to explain the origin of the spiral patterns from a rotating BEC as interference fringes from matterwave emitters with different momenta, see text. (b) Correlation functions $g^{(2)}_t$ in polar coordinates $(r,\,\phi)$ for $t=40$~ms image in Fig.~\ref{fig:vortexr}(a). Red dashed lines show linear fits to the
correlations between $r$ and $\phi$.(c) Effective angular velocity $\omega_e$, expressed in units of $\hbar/mR^2$, for condensates with different winding number $l_0$. Blue circles are from simulations and red circles are from experiments. Error bars represent 1-$\sigma$ standard deviation.
}
\label{fig:vortexAna}
\end{figure}

Our simulations and experiments show a consistent picture that the jet emission displays a spiral pattern in the presence of vorticity in the BEC. When the winding number is positive, the spirals are clockwise. The spiral emission pattern is the key observable that determines the winding number of the condensate.

This spiral pattern can be understood based on a semi-classical picture. Considering atoms inside the rotating condensate as independent emitters, an atom has a unique momentum $\bf{k}$ of magnitude $l_0/ r$ along the transverse direction. When two such atoms collide inelastically, they are excited to new momenta ${\bf{k}}\pm {\bf{k_f}}$, where $|{\bf{k_f}}|= k_f$. For an observation point outside the sample, jets emitted from different parts (``sources") of the condensate overlap and interfere, and the observed spirals are the resulting interference fringes.

To see the connection between the direction of the spiral and the angular momentum, we note that when the observer moves away from the condensate, the phase of the matterwave with relatively large momentum accumulates faster. Thus the fringe curves toward the jet with the higher momentum, namely, $\bf{k}+\bf{k_f}$, to maintain the same interference condition, see Fig.~\ref{fig:vortexAna}(a). Theoretical analysis suggests $d\phi/dr= -\eta l_0/(k_fR^2)$ \cite{Supple} with $\eta$ being a dimensionless constant. This equation describes the observed spirals.

To test these predictions, we evaluate the correlation function between two points with radial distance $r$ and angular distance $\phi$, namely,
\begin{equation}
    \label{eq:g2arg}
 g_t^{(2)}=\frac{\int d\phi^\prime dr^\prime \langle n(r^\prime,\phi^\prime)n(r^\prime+r, \phi^\prime+\phi)\rangle}{2\pi L_0\tilde{n}^2},
 \end{equation}
 where the integration of $r^\prime$ covers the interval $L_0$ that jets manifest \footnote{Assuming jets appear within $r_{min}<r<r_{max}$, we integrate $r^\prime$ in the range such that both measurement points at $r^\prime$ and $r^\prime+r$ are within this ring area.}  and $\tilde{n}$ is the mean density in the interval.

The spiral pattern associated with the jet substructure can be understood as representing a linear relation between the radial and angular
distances in the emission.
See Fig.~\ref{fig:vortexAna}(b), where the red dashed lines show linear fits to the
correlations involving $r$ and $\phi$.
This linear dependence suggests that the emission emerges with an effective angular velocity $\omega_e=-(\hbar k_f/m)d\phi/dr$, which can be compared with the winding number of the condensate according to
\begin{equation}
\label{eq:omega}
\omega_e=\eta\frac{l_0\hbar}{mR^2},
\end{equation}
see Fig.~4(c). From simulations, we determine $\eta= 2.90$ for $l_0=\pm 1$ and $\eta= 2.19$ for $l_0=\pm 2$. We speculate that the decrease of $\eta$ for larger $|l_0|$ is a result of the instability of a vortex-containing-BEC with $l_0=\pm 2$. A vortex with $l_0=2$ will quickly decay into two vortices with $l_0=1$, and the finite spatial separation between them reduces the effective angular velocity. For a classical, rigid uniform disk with the same radius $R$, we expect that the angular velocity is $\omega_e=\eta_{cl} l_0\hbar/(mR^2)$ with $\eta_{cl}=2$.

The same analysis on the experimental data also yields a linear relationship between $r$ and $\phi$ in the correlation function. Based on multiple repeated experiments, we find that $\eta l_0$ takes on quantized values of $\eta l_0=-3.07(3)$, $-0.10(6)$ and $3.0(1)$, which are in very good agreement with the simulation results for $l_0=-1$, $0$ and $1$, see Fig.~\ref{fig:vortexAna}(c). The agreement between experiments and simulations confirms our scheme to reveal the helicity of a BEC directly from the jet emission pattern.

In conclusion, we show in two examples that jet sub-structure,
also of interest in particle physics \cite{RMP91_045003,RMP90_025005}, is a powerful tool to probe the wavefunction of the condensate. In particular, topological defects like solitons and vortices can be readily identified from the jet correlations.
These two-body correlation functions
$g^{(2)} $   can be seen to be directly determined by
the function $F$ introduced earlier \cite{Supple}. This function, in turn
enables us to arrive at essential information about the phase
and density in a condensate.
As is consistent with theoretical expectations, we find
excellent agreement between our experiments and simulations.

We acknowledge Miguel Arratia, S.~Fnu and X.~Wang for helpful discussions, and I.~Aronson and A.~Glatz for the numerical code. We acknowledge support by the U.S. Department of Energy, Office of Basic Energy Sciences, under contract number DE-SC0019216, the Army Research Office under Grant No. W911NF-15-1-0113, and the University of Chicago Materials Research Science and Engineering Center, funded by the National Science Foundation under Grant No. DMR-1420709. L. Feng acknowledges support from the MRSEC Graduate Research Fellowship.

%

%\pagebreak
%\newpage
\clearpage
\widetext
\begin{center}
\textbf{\large Supplement: Jet Sub-structure in Fireworks Emission from Non-uniform Bose-Einstein Condensates}
\end{center}

\beginsupplement

In this supplement, we first present details of the experimental procedure and data analysis for our study on jet emission from both split and vortex-embedded BECs. We then provide a theory for inverting the complete correlation functions to recover the general initial wavefunctions. Lastly, more quantitative derivations and supplementary simulation results are presented.

\section{1. Experimental details of study on jet emission by split BECs}
\emph{Experimental procedure-} We start with 3D BECs of 60,000 cesium atoms loaded into an elliptical crossed dipole trap. Then 40,000 atoms are adiabatically transferred within 760~ms into a disk-shaped dipole trap with a diameter of 18~$\mu m$ in the horizontal direction and a 6-$\mu m$ wide central barrier along the diameter that splits the BEC into two halves. The potential barriers are provided by a blue-detuned laser at 788 nm. The laser beam profile is shaped by a digital micromirror device (DMD) and projected to the atom plane through a high-resolution objective (of 1 micron resolution). The resulting circular potential well and the central barrier have barrier height of $h\times 140~\mathrm{Hz}$ and $h\times 42~\mathrm{Hz}$, respectively. Atoms are tightly confined in the vertical direction with a $1/e^2$ radius of 0.8~$\mu m$ and a harmonic trap frequency of 259~$\mathrm{Hz}$. The phase coherence of the two half BECs is maintained, which is revealed by the interference fringes formed during time-of-flight.

Then we use a DMD to project a 788~nm light pulse of duration $\tau = 0.4~\mathrm{ms}$ on one half of the BEC to induce a relative phase shift. The imprinted phase $\theta_s = -V_s\tau/\hbar$ is tuned by changing the light pulse intensity that determines the light shift $V_s$. The potential gradient in the imprinting process applies force to the imprinted half and `kicks' it away from the unimprinted half. The small envelope shift between $\theta_s=-\pi$ and 0 in the experiment in Fig. \ref{fig:artrap}(c) comes from this "kick" effect. About 1 ms after the phase imprinting, we apply an oscillating magnetic field in the vicinity of a Feshbach resonance to the BECs, which modulates the atomic s-wave scattering length as $a(t) = a_{dc} + a_{ac}\sin(\omega t)$ with a small offset $a_{dc} = 9a_0$ and a large amplitude $a_{ac} = 47a_0$, at frequency $\omega/2\pi = 2.1~\mathrm{kHz}$. After modulating the interaction for 34~ms, we perform \textit{in situ} absorption imaging through the same high-resolution objective and a CCD camera to observe the structure of ejected atomic jets. For measuring the connected correlation $g^{(2)}_{\mathrm{conn}}(\phi)$, we wait for an additional 10 ms after the 34~ms interaction modulation before performing imaging. We do this because when the jets fly to the far field, the oscillation of the $\pi-$peak in the correlation function becomes prominent.

\begin{figure}[h]
\includegraphics[width = 0.5\textwidth]{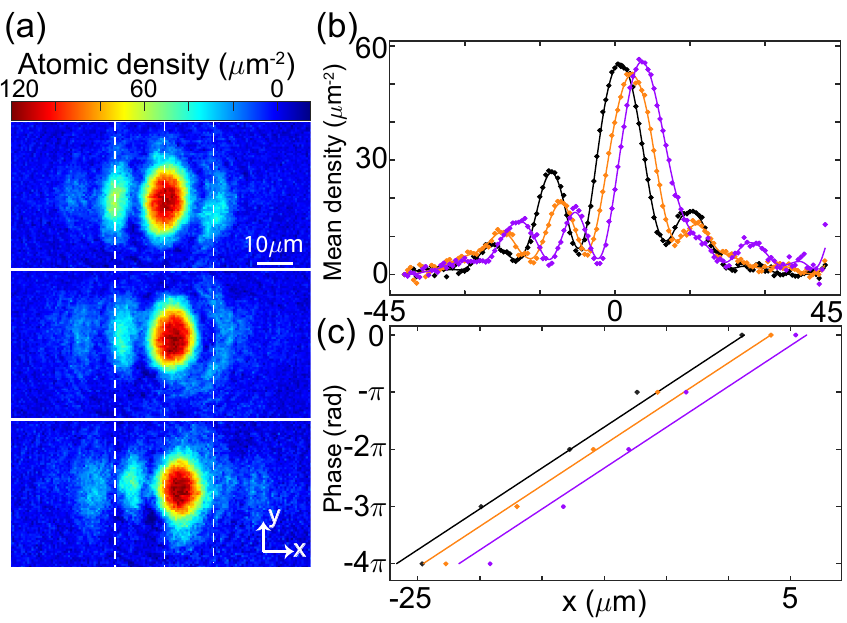}
\caption{Calibration of the imprinted relative phase between two halves of the BEC through TOF imaging. (a) Single-shot 30-ms TOF imaging of the interference fringes for imprinted phase $\theta_s$ around 0 (top), $-\pi/2$ (middle) and $-\pi$ (bottom). (b) Mean atomic density along $x$ direction corresponding to single shot images in (a) for imprinted phase $\theta_s$ around 0 (black), -$\pi/2$ (orange) and -$\pi$ (purple). (c) Peak and valley positions corresponding to the mean atomic density distributions in (b) with the same color scheme versus their phase. The phases are assigned according to whether they correspond to a peak (even multiples of $\pi$) or a valley (odd multiples of $\pi$). }
\label{fig:Supp_Fig2}
\end{figure}

\emph{Imprinted phase calibration-} In order to measure how much phase is imprinted onto one half of the BECs through the short light pulse, we let the two half BECs expand freely for 30 ms right after the phase imprinting. The two parts of the condensate acquire momentum $\pm k_t$ after being released and form interference fringes when they overlap in space. In this way, the phase shift of the fringes as shown in Fig.~\ref{fig:Supp_Fig2}(a) for different light pulse intensity can reflect the value of $\theta_s$. In Fig.~\ref{fig:Supp_Fig2}(b), the corresponding mean atomic density distributions along the $x$ direction $n(x) = A(x)[\cos(k_tx+\theta_s)+C]$ are shown (the origin of coordinates is set according to the no-phase-imprinting case where $\theta_s = 0$.). We identify the positions of density peaks and valleys on the left of the highest peak and assign a phase of either even or odd multiple of $\pi$ as shown in Fig.~\ref{fig:Supp_Fig2}(c). The data are fit linearly and the change of $y$-intercept corresponds to the change of imprinted phase. The black curve is when no light pulse is applied and serves as a reference at $\theta_s=0$. By comparing the $y$-intercept to it, the orange and purple curves yield the values of $\theta_s$ that are near $-\pi/2$ and $-\pi$ as in Fig.~\ref{fig:artrap}(d).

\section{2. Details of extracting phase $\delta\theta$ from fitting correlation functions near $\phi=\pi$}

To understand the interference pattern of split BECs we refer to the
double-slit interference model and make an approximate analogy between
the split BEC and a conventional double slit problem. In this model the
far field (Fraunhofer) diffraction intensity is proportional to
$$\textrm{sinc}^2\left(\frac{\pi W\sin\alpha}{\lambda}\right)\cos^2\left(\frac{\pi D\sin\alpha}{\lambda}-\theta_r/2\right),$$
where $\alpha$ is the diffraction angle, $\lambda$ is the light wavelength, $D$ is the distance between the slit centers, $W$ is the width of each slit, and $\theta_r$ is the relative phase between the light beams that pass the two slits.

We fit the oscillatory correlation function $g^{(2)}_{conn}(\phi)$ near
$\phi=\pi$ with the following function to extract the relative phase
$\delta\theta$ between the two half BECs, see Fig.~\ref{fig:artrap}(c):
$$f(\phi) = A\,\textrm{sinc}^2[b(\phi-\pi-c)]\left[\cos^2\left(\frac{k\phi-k\pi+\delta\theta}{2}\right)+d\right]+f_0,$$
where the sinc envelope captures the finite size of each half BEC, the
cosine term describes the matterwave interference fringes, $b$ describes
the envelope oscillation frequency, $c$ accounts for the small center of
mass motion, $k$ depends on the separation between the two half BECs,
and the parameters $d$ and $f_0$ describe the offset of the fringes and
the envelope function, respectively. By fitting the $\pi$ peaks using
this functional form, we extract the relative phases $\delta\theta$
shown in Fig. \ref{fig:artrap}(d).

In our experiments we determine $k$ by fitting the
correlation function $g^{(2)}_{conn}(\phi)$ for condensates with no phase imprinting. The
fitted value of $k$ is then fixed for other situations with nonzero
imprinted phase. Since our samples form 2 semi-circles instead of 2
slits, we do not expect the sinc function to precisely describe the
measured envelope function near the $\pi$ peaks. We have verified that the phase
shifts $\delta \theta$ we extracted have negligible dependence on the form
of the envelope function and offsets.

\section{3. Analysis for the experimental data on jet emission by vortex-embedded BECs}
\begin{figure}
    \centering
    \includegraphics[width = 1\textwidth]{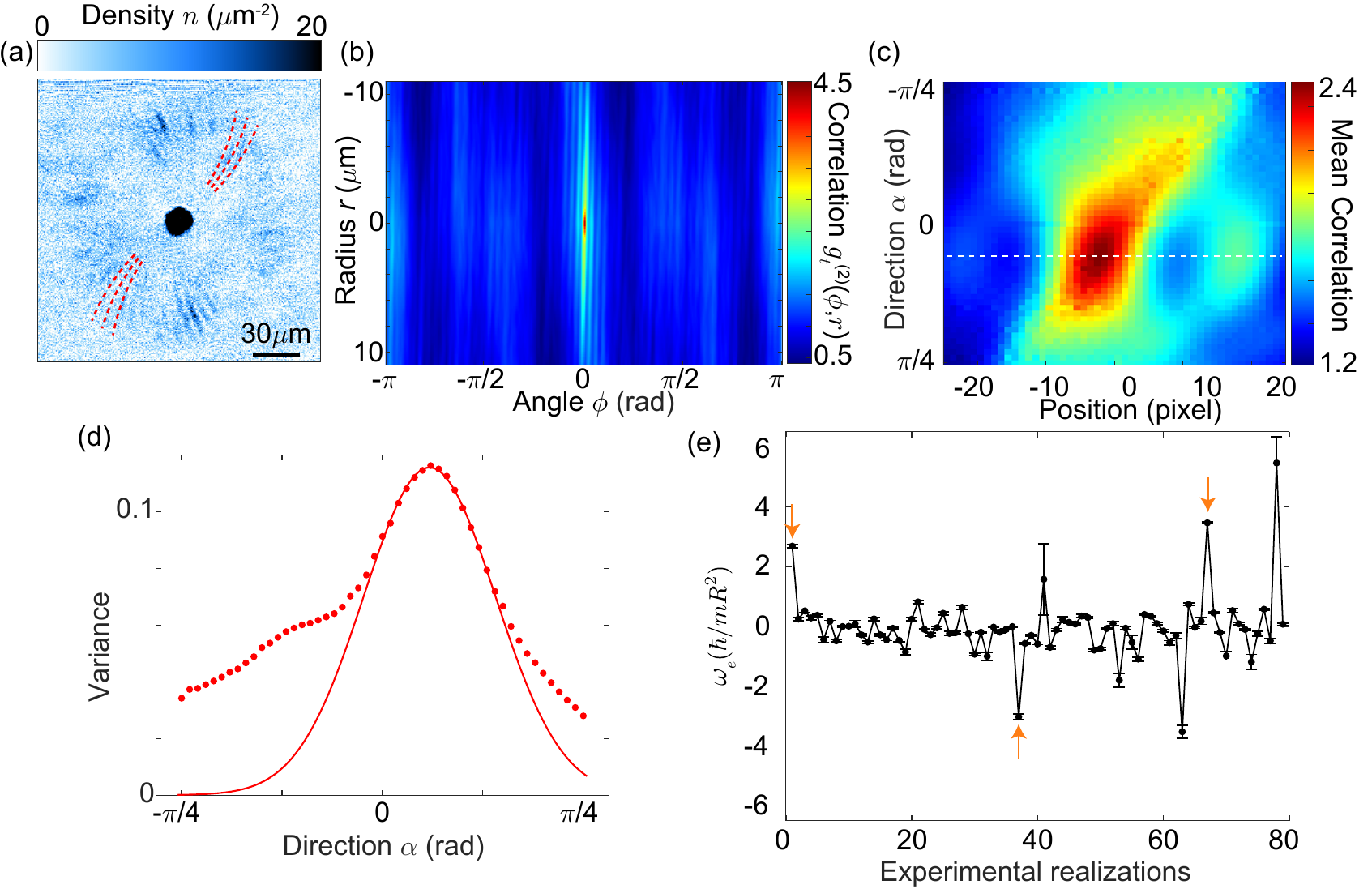}
    \caption{Determination of vortex winding number from emission patterns of vortex-embedded BECs. (a) Example emission pattern from a vortex-embedded BEC. (b) Auto-correlation of atomic density of emitted jets in (a) as a function of relative displacement $r$ and $\phi$ in radial and azimuthal directions, respectively. (c) Distribution of mean correlation averaged over different directions in (b) for the lower half with positive r. The values for the central 50 pixels are shown here. The white dashed line indicates the direction where the variance of mean correlation reaches maximum. (d) The data points are variance of the mean correlation distribution averaged from different directions in (b). The solid line is a Gaussian fit for the central 13 data points around the maximum, which determines the peak position precisely. The direction where the maximum variance occurs corresponds to the slope $d\phi/dr$ of the fringes around $\phi = 0$ in (b), which then corresponds to the curvature of spirals in (a). (e) Angular velocity $\omega_e = d\phi/dr\times k_fR^2$ calculated from the slope $d\phi/dr$ following the procedure in (a)-(d) (corresponding to the $1^{st}$ experimental realization). The calculation is based on unsorted experimental realizations. The orange arrows indicate the data points identified as $l_0 = \pm 1$ with small error bars $<0.2$. Other data points with error bars less than 0.2 are averaged and are identified as $l_0 = 0$. Data points with larger error bars are ignored. The results are shown as red circles in Fig.~\ref{fig:vortexAna}(c). The error bars are 1-$\sigma$ standard deviation.}
    \label{fig:FigS5}
\end{figure}

In order to extract the vortex winding number from emission patterns of vortex-embedded BECs in experiments, we first calculate the auto-correlation of atomic density in emitted jets $g_t^{(2)}(\phi,r)$ as a function of azimuthal and radial displacements $\phi$ and $r$, as given by Eq. \eqref{eq:g2arg}.
\begin{comment}
\begin{align}
    g_t^{(2)}(\phi,r) = \frac{\langle n(r_1,\phi_1)n(r_1+r,\phi_1+\phi)\rangle}{\langle n(r_1,\phi_1)\rangle\langle n(r_1+r,\phi_1+\phi)\rangle},
    \label{g_t}
\end{align}
where $\langle\cdot\rangle$ means averaging over all possible positions $(r_1,\phi_1)$ in the region of atomic jets.\end{comment}
One example of the correlation function $g_t^{(2)}(r,\phi)$ is shown in Fig. \ref{fig:FigS5}(b) for the emission pattern from experiment in Fig. \ref{fig:FigS5}(a). It can be seen that there are fringes near $\phi = 0$ in the auto-correlation with non-zero slope $d\phi/dr$, which is proportional to the vortex winding number. Next, we quantitatively extract the slope from the pattern of fringes in Fig. \ref{fig:FigS5}(b) using the pattern recognition algorithm which is described in the next paragraph, thus enabling us to extract the winding number.

To recognize the fringes, we can average the two-dimensional correlation function $g_t^{(2)}(r,\phi)$ along different directions. The direction along which the
mean correlation shows the oscillation structure most clearly, corresponds to the slope of those fringes. The mean correlation distribution in the central region of 50 pixels at different directions with angle $\alpha$ is shown in Fig. \ref{fig:FigS5}(c), where $\alpha = 0$ is along the negative $r$ axis in Fig. \ref{fig:FigS5}(b) and $\alpha = \pi/2$ is along the positive $\phi$ axis. We use the variance of the mean correlation distribution in certain directions to characterize its contrast, which is shown in Fig. \ref{fig:FigS5}(d). There is a clear peak in the variance and we use a Gaussian function to fit the 13 data points around the maximum to find the peak position $\alpha_p = 0.184 ~\mathrm{rad}$ and use the uncertainty of the fit as error bars. Then the slope $d\phi/dr = \chi\tan(\alpha_p)$ is determined, where $\chi = 0.0216~\mathrm{rad/\mu m}$ is the ratio between the resolution in angular and radial direction in Fig. \ref{fig:FigS5}(b). Finally, the angular velocity $\omega_e = d\phi/dr\times k_fR^2$ in units of $\hbar/mR^2$ is obtained.

We apply the same procedure as above for images from 79 repetitive experimental realizations and obtain their angular velocity $\omega_e$ as shown in Fig. \ref{fig:FigS5}(e). The emission pattern in (a) corresponds to the first data point in (e) and the data points indicated by the orange arrows are the experimental data points shown in Fig. \ref{fig:vortexAna}(c). We determine the vortex winding number $l_0$ by comparing the measured angular velocity $\omega_e$ to the corresponding simulation results. In addition, we see that most of the measurements have zero winding number, since the vortices are non-deterministically generated.

\section{4. Inverting theory using complete correlation functions}
In this section, we show that if one has full knowledge of all the two-operator correlation functions: $\langle a_i a_j\rangle,\,\langle a_i a_j^\dagger\rangle,\,\langle a_i^\dagger a_j\rangle,\,\langle a_i^\dagger a_j^\dagger \rangle $ at a given time $t$, one can recover the initial wavefunction.

In the interaction picture and dropping the far off-resonant terms, we obtain
an effective Hamiltonian
as
\begin{equation}\label{eq:interact}
H=\frac{1}{2}\sum_{i,j}P_{i j}a_i a_j+ h.c.
\end{equation}
where $P_{ij}\approx U_1F^*(i,j)/2i$. Assuming we have $N$ bosonic modes $a_1,\,a_2,\ldots,a_N$, we define the vector $A=(a_1,a_1^\dagger,\ldots,a_N, a_N^\dagger)^T=(A_1,A_2,\ldots,A_{2N-1},A_{2N})^T$. Then any matrix operating on $A$ that has $2N\times 2N$ dimension can be reduced to a tensor product between a $N\times N$ matrix and a $2\times 2$ Pauli matrix. From Eq. \eqref{eq:interact}, we arrive at
\begin{equation}
i\hbar \frac{dA}{dt}=[A,H]=i(\textrm{Re }P\otimes \sigma_y-\textrm{Im } P\otimes\sigma_x)A=iKA
\end{equation}
where $\textrm{Re }P$ and $\textrm{Im }P$ are respectively the real and imaginary parts of the matrix $P$.
Then, we have $A(t)=e^{Kt/\hbar} A(0)$. If we define the matrix $\gamma $ as
\begin{equation}
\gamma_{ij}=\langle A_iA_j^\dagger+A_j^\dagger A_i\rangle,
\end{equation}
we then find $\gamma(t)=e^{Kt/\hbar} \gamma(0)e^{Kt/\hbar}=e^{2Kt/\hbar}$ where $\gamma(0)$ can be easily derived to be the identity matrix as we start from vacuum. $\gamma(t)$ here is simply composed of all the two-operator correlation functions at $t$. Therefore, from the correlation functions, we can extract $\gamma(t)$ and characterize $P$ as
\begin{equation}
\begin{aligned}
K=&\frac{\hbar}{2t}\ln \gamma(t)\\
\textrm{Re } P=&\frac{1}{2}\textrm{Tr}_\sigma \left[K\times (I\otimes\sigma_y)\right]\\
\textrm{Im } P=&-\frac{1}{2}\textrm{Tr}_\sigma \left[K\times (I\otimes\sigma_x)\right].
\end{aligned}
\end{equation}
where $I$ is the $N\times N$ identity matrix, Tr$_\sigma[\ldots]$
involves the trace over the $2\times 2$ dimension. Since $F(i,i^\prime)$ can be obtained from the $P$ matrix, and the original wavefunction $\psi_0$ can be derived from $F(i,i^\prime)$ according to Eq. \eqref{eq:funcF} up to a sign uncertainty, we then can recover the wavefunction,  $\psi_0$.

\section{5. Results and analysis for soliton-embedded BECs}

\begin{figure}[h]
\includegraphics[width=0.5\textwidth]
{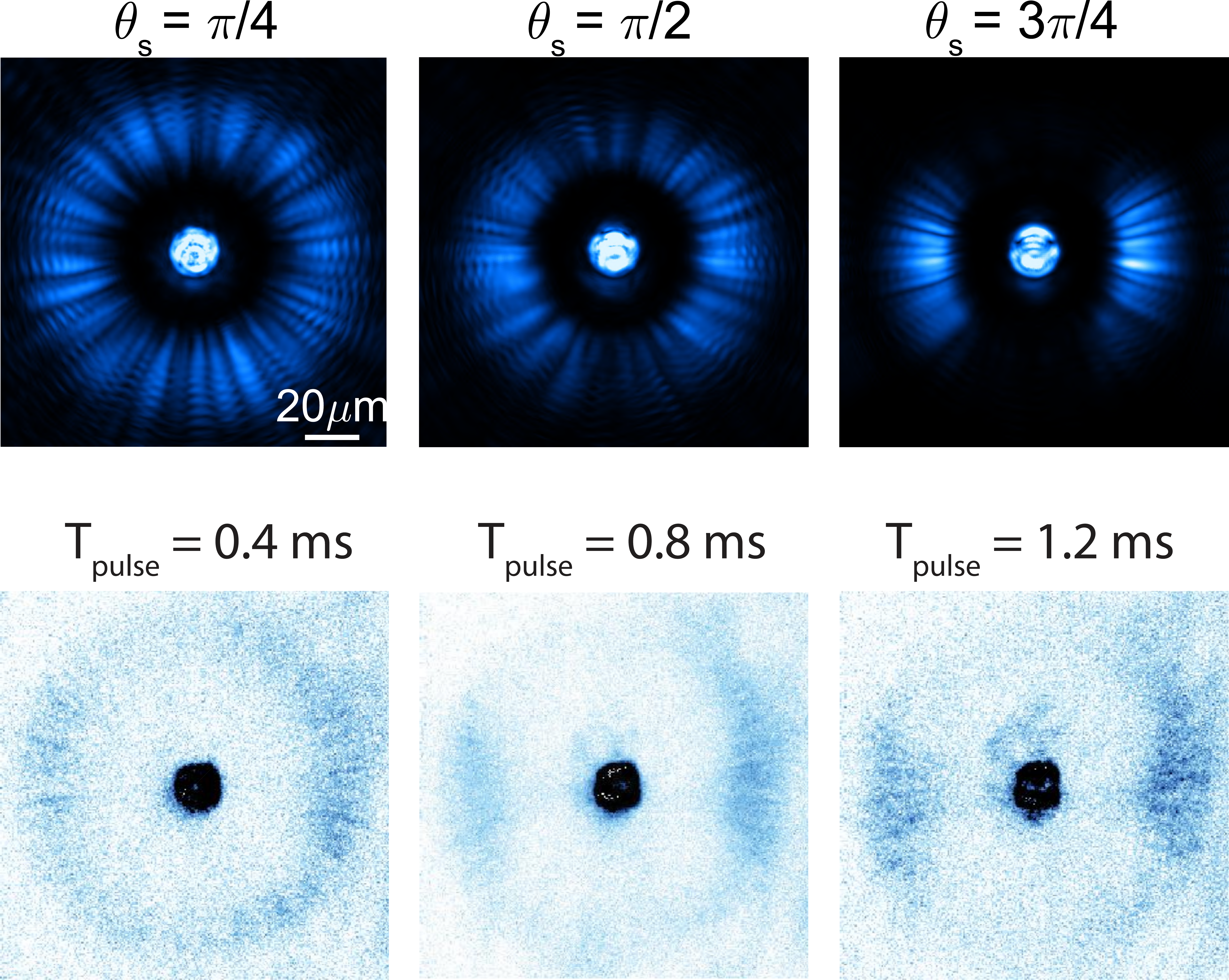}
\caption{Emission pattern for condensates with lower half phase-shifted relative to the upper. Upper panel shows results from GP simulations with the imprinted phase $\theta_s$ as $\theta_s=\pi/4,\,\pi/2,\,3\pi/4$ from left to right. Lower panel presents experimental data of phase-imprinted condensates. As the pulsing time increases, $\theta_s$ grows towards $\pi$ (however, there is no exact calibration of the phase so the experimental $\theta_s$ values are not the same as the ones in the upper simulations panel). In both simulations and experiments, one can see that the resulting soliton formed along the horizontal axis
yields the strong directionality of the emission pattern.}
\label{fig:S_soliton}
\end{figure}
Solitons can naturally  arise from phase imprinted condensates (which are not split by central barriers). There is then a notable directionality in the
stimulation process with the resulting emission highly oriented along the direction of the
phase slip boundary shown in Fig. \ref{fig:S_soliton}.
This directionality for different phases $\theta_s$ primarily results from
the different level of density depletion for different solitons. The closer the imprinted
phase $\theta_s$ is to $\pi$, the stronger the density depletion is. Therefore, for excitations
propagating in the vertical directions, there are fewer atoms along their path giving rise to less stimulation.
Experimentally, the effect of varying $\theta_s$ is studied and is qualitatively consistent with theory,
see lower panel in Fig. \ref{fig:S_soliton}.

\begin{figure}[h]
\includegraphics[width=0.6\textwidth]
{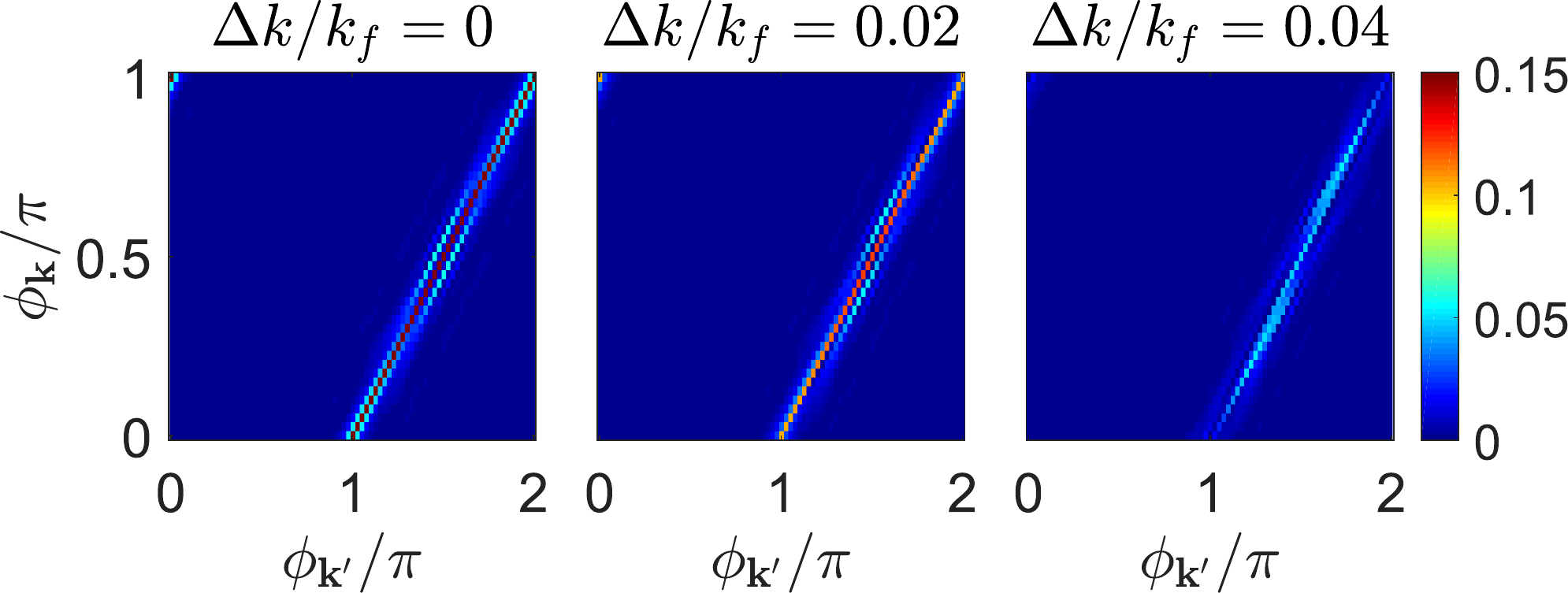}
\caption{Numerical calculation of the absolute value of the pair function $|F|$ for different $k$. Since the system is symmetric, we restrict the direction of $\bf{k}$ to the $[0,\,\pi]$ domain. The horizontal axis is the azimuthal angle of $\bf{k^\prime}$ while the vertical axis is the angle of $\bf{k}$. The corresponding $|F|$ is denoted by the color value. From left to right, $\Delta k/k_f=0,\,0.02,\,0.04 $, and the preferred emission direction (associated with momentum modes that yield large $|F|$) is near $\phi_{\bf{k}}=\pi/2$ (which is the horizontal direction as consistent with the assumptions in
our calculations). }
\label{fig:int}
\end{figure}

\begin{figure}[h]
\includegraphics[width=.5\textwidth]
{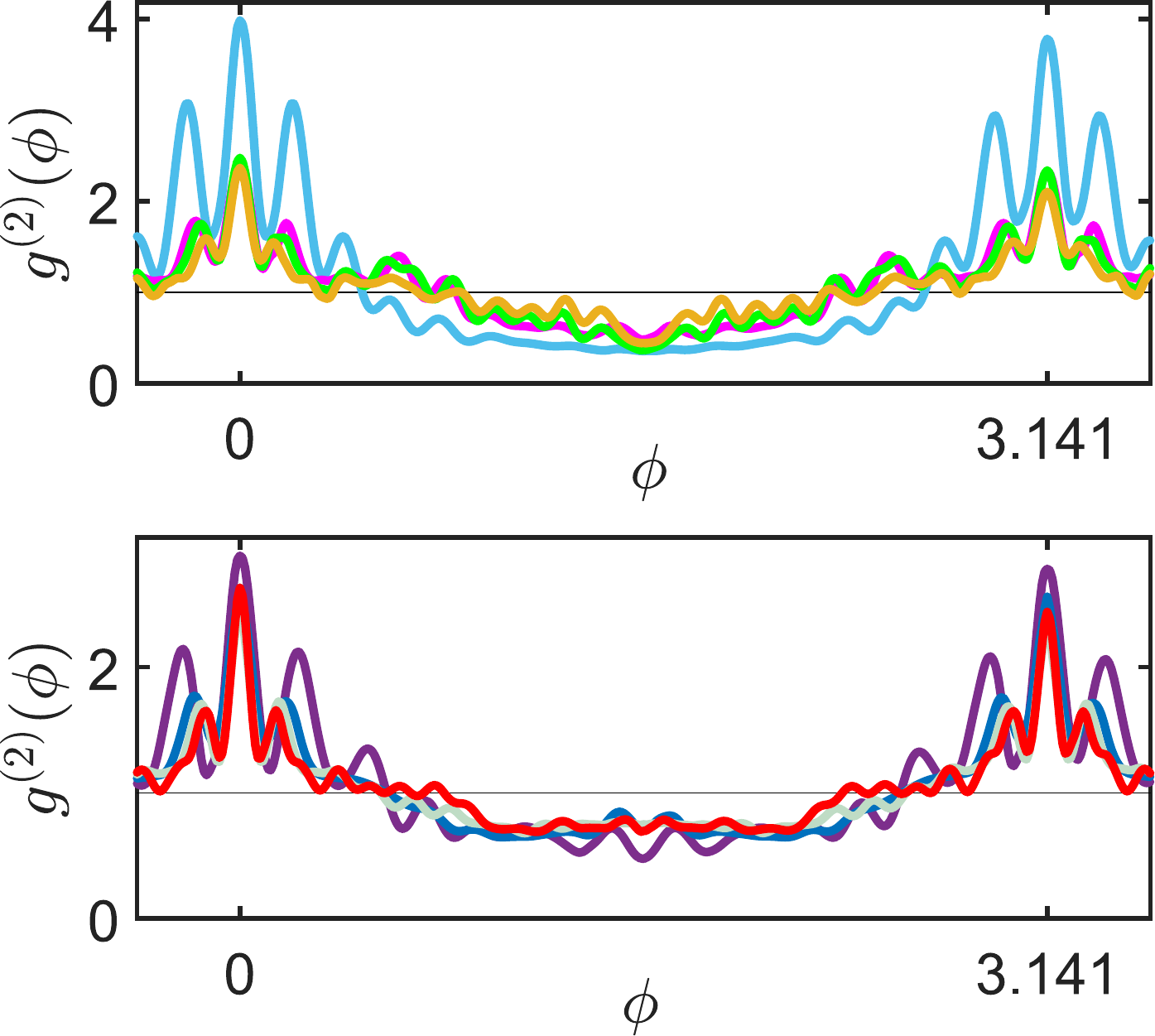}
\caption{Real-space density correlation function $g^{(2)}(\phi)$ taken at distance $\sim70\,\mu$m from the trap for oriented fireworks in the dark soliton case. The double-slit interference causes clear oscillations near the $0,\pi$ peaks. (a) Correlation function for different trap radii $R$ at $\omega=$2000 Hz. The solid lines correspond to $R=6$ (blue), 7 (magenta), 8 (green), and 9 $\mu$m (orange), respectively. (b) Correlation function for various modulation frequencies at $R=7\,\mu$m. The solid lines correspond to $\omega=1500$ (purple), 2000 (blue), 2500 (green), and 3000 (red) hz, respectively. One can see that the oscillations near the peaks become faster with increased $\omega$ or $R$.}
\label{fig:corr}
\end{figure}

Let us now understand this more quantitatively, taking the dark soliton ($\theta_s=\pi$) case as an example. As explained in the Introduction, the parametric amplification is determined by the pair function $F$.
A numerical calculation of $F$ for the dark soliton wavefunction
\begin{equation}
\label{eq:dark}
\psi_0(r,\phi)
=\sqrt{n_0}\tanh\left(\frac{y}{\sqrt{2}\xi}\right) \end{equation}
is presented in Fig. \ref{fig:int}. This is done in a plane wave basis, describing the overlap between a pair of ${\bf{k}},\,{\bf{k^\prime}}$ modes with the original state. The corresponding equation of motion for the mode in the rotating wave approximation is then
\begin{equation}\label{eq:eom_S}
\begin{aligned}
i&\hbar \frac{\partial a_{\bf{k}}(t)}{\partial t}-(\frac{\hbar^2k^2}{2m}+\mu)a_{\bf{k}}(t)\\
\approx& -\frac{U_1e^{-i\omega t}}{2i}\int d{\bf{k^\prime}}a_{\bf{k^\prime}}^\dagger(t)F({\bf{k}},{\bf{k^\prime}}).\\
\end{aligned}
\end{equation}
In Fig. \ref{fig:int}, to respect energy conservation, we take ${k^\prime}^2+k^2\simeq 2k_f^2$ where $k_f = \sqrt{m\omega/\hbar}$. One can see, as $\Delta k=k-k_f$ increases, or when including a finite angular uncertainty leads to
a nonzero metric ($d{\bf{k^\prime}}\neq0$), the integral peaks at certain $\phi_{\bf{k}}$ values, giving rise to the oriented emission. In all cases, the peak value occurs at $\phi_{\bf{k^\prime}}\approx \phi_{\bf{k}}+\pi$, leading to the perfect $0-\pi$ inversion symmetry in the jet pair production process, as consistent with the double slit interference in the $\theta_s=\pi$ case.
There are other behaviors similar to the double-slit interference. When one increases the trap radius $R$, the effective distance between the two slits ($\sim R$) also increases, giving rise to a faster oscillation of the interference fringes. Also, by increasing the modulation frequency, the effective wavelength of the excitations decreases, leading to a faster oscillation as well, see Fig. \ref{fig:corr}.

One might wonder how this compares with the simple, uniform (in both phase and density) condensate picture which is asymptotically approached at vanishing $\xi$. This is analogous to the crossover between double-slit and single-slit experiments. As mentioned in Sec. 1, for double slits, the far-field (Fraunhofer) diffraction intensity is $\propto \cos^2\left(\pi D\sin\alpha/\lambda\right) \textrm{sinc}^2\left(\pi W\sin\alpha/\lambda\right)$, where $\alpha$ is the diffraction angle, $\lambda$ is the light wavelength, $D$ is the distance between the slit centers and $W$ is the width of each slit. For the single slit, the diffraction is $\propto \textrm{sinc}^2\left(\pi L\sin\alpha/\lambda\right)$, where $L$ is the width of the single slit. When the separation between the two slits approaches zero, $D=W=L/2$, the double-slit diffraction intensity acquires the same expression as that of the single slit. The additional envelope $\textrm{sinc}\left(\pi W\sin\alpha/\lambda\right)$ assists the smooth crossover. For the soliton case, there is a similar distance $D$ between the centers of two separated parts. When $\xi$ is very small, $D$ is solely determined by the trap radius $R$. The frequency of the fast oscillation becomes a constant. But the additional envelope evolves, decreasing faster and leaving only the central peaks near $\phi=0$, $\pi $
which are not destroyed. This approaches the single-peak picture of correlation functions for the usual uniform condensate.

\section{6. Results and analysis for vortex-embedded BECs}

\begin{figure}[h]
\includegraphics[width=.5\textwidth]
{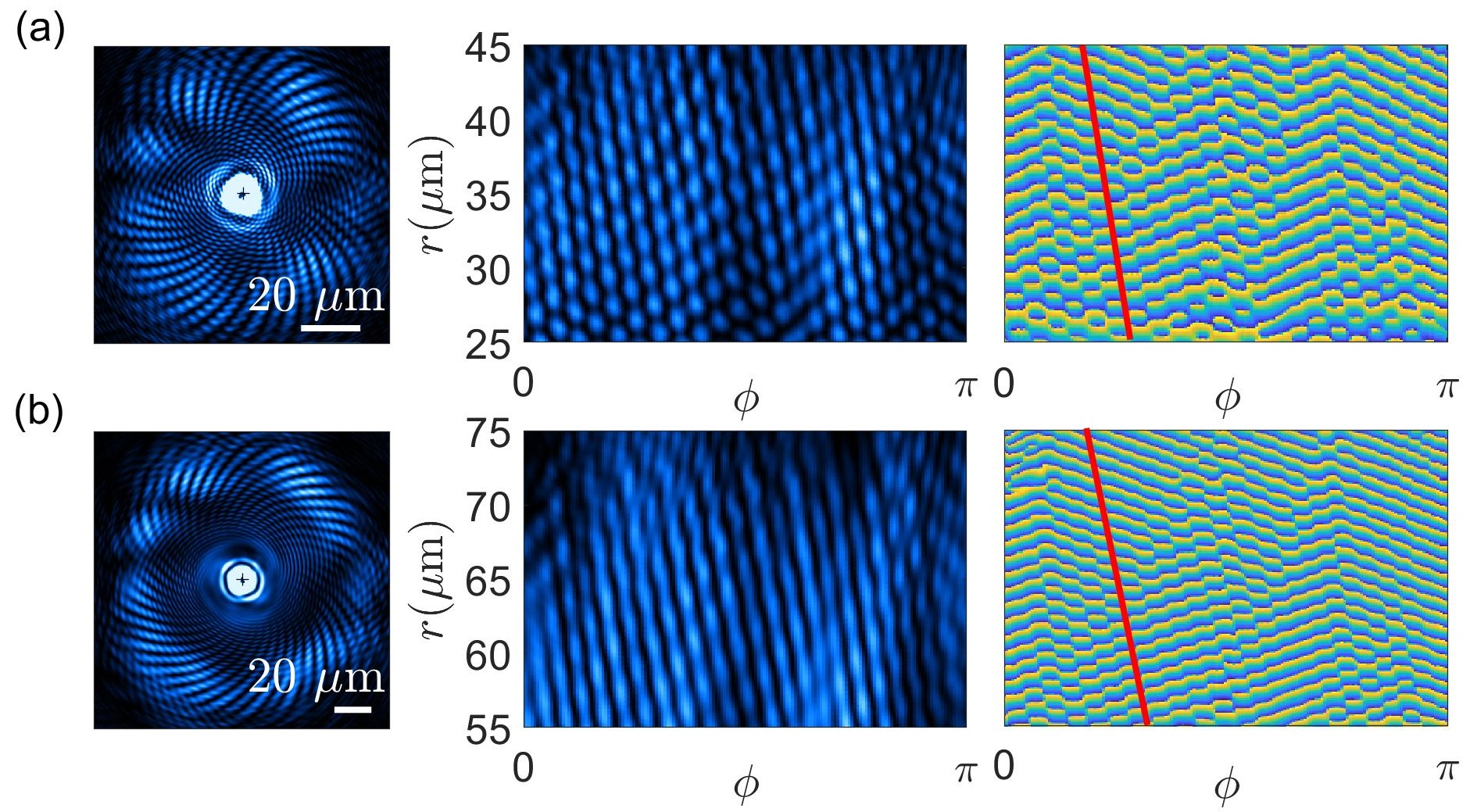}
\caption{Real-space images for spiral jets emitted by modulated condensate of $l_0=2$ from numerical simulations.
From left to right: density image in Cartesian coordinates, density image in polar coordinates, and phase image in polar coordinates. (a) Plotted at $t$=30 ms. (b) Plotted at $t=40$ ms. The near field pattern (a) has additional fringes on top of the spirals. This is caused by the interference from higher harmonics ($k/k_f=\sqrt{2},\,\sqrt{3},$ etc.). Since different $k$ modes quickly separate at larger $t$, in (b) we see the diffractive fringes are substantially reduced.
The red lines are a guide-to-the-eye indicating the slope of the spiral trajectories.}
\label{fig:TOF}
\end{figure}

Let us now look at the time of flight images at different times for the vortex-embedded BEC. In Fig. \ref{fig:TOF}, we show the early time ($t=30$ ms) and late time ($t=40$ ms) emission pattern in the different coordinates. We can see that the trajectories are real spirals which convert to straight lines in polar coordinates. And the slope in polar coordinates is a constant throughout the time of flight as determined only by the trap and driving parameters. As we look at phase maps, we immediately notice that the phase accumulates not only in the radial direction but also in the angular direction. And its angular phase winding direction can randomly be positive or negative. This reveals the physical mechanism behind this spiral emission pattern: two-mode interference.

\begin{figure}[h]
\includegraphics[width=0.5\textwidth]
{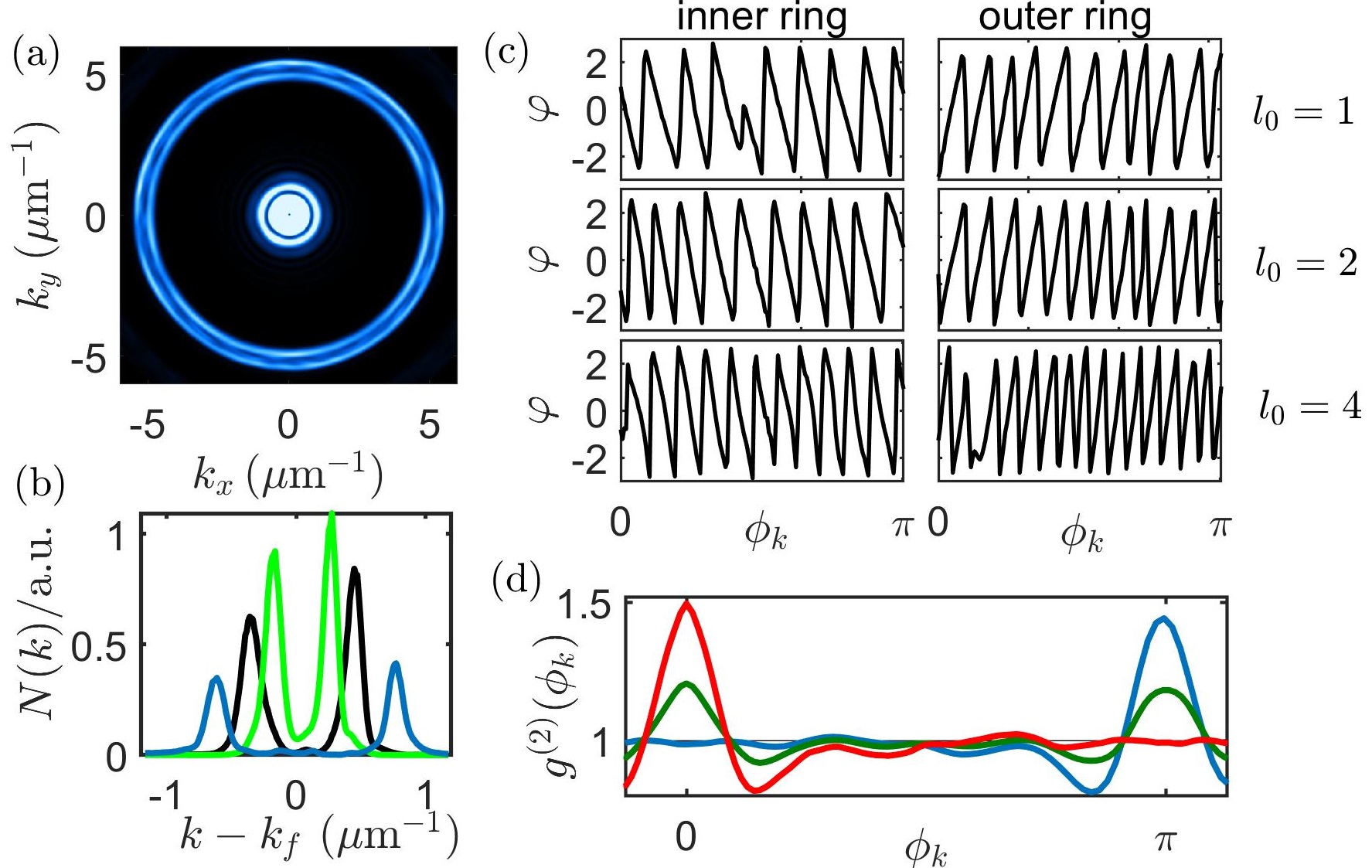}
\caption{$k$-space analysis of emission pattern by vortex-embedded condensate from GP simulations.
(a) Density image in Cartesian coordinates for $l_0=1$, which clearly shows two dominant $k$ values of the parametrically amplified modes. (b) Distribution of emitted (excited) bosons at different wavenumber $k$: $N(k)=\int d\phi N(\vec{k}) k$.
The solid lines, respectively, correspond to the $l_0=1$ (green), 2 (black), and 4 (blue). (c) Phases of the modes in the two major momentum rings as a function of the mode direction $\vec{k}$ for $l_0$=1, 2, and 4 from top to bottom. (d) Correlation function for spiral fireworks in the vortex case.
The solid lines correspond to intra- (red), inter- (blue), and total two-ring (green) correlation functions.}
\label{fig:vortexk}
\end{figure}

Looking at the $k$-space distribution of the excitations, one can see that the momentum values peak at two values instead of one. Each of them also is associated with a particular ``angular momentum": a fast phase winding along the azimuthal direction, see Fig. \ref{fig:vortexk}. Therefore, in real space, there are two modes of different radial momenta and angular momenta interfering. This, as we explain in more detail in the following subsection, gives the spiral trajectories.

\subsection{6.1. Analytical and numerical derivations of the two-ring structure}

Since the trap respects cylindrical symmetry, it is more convenient to use the angular momentum eigenstates as the basis: $\varphi_{lk}=\sqrt{k/2\pi}e^{il\phi}J_l(kr)$ where $J_l$ is the Bessel function with index $l$. Using Eq. \eqref{eq:vortex}, Eq.
\eqref{eq:funcF} is then rewritten as
\begin{equation}\label{eq:vortexF}
F(l, l^\prime; k,k^\prime)=\sqrt{kk^\prime}\int_0^R
rdr J_{l_0+\Delta l}(kr)J_{l_0-\Delta l}(k^\prime r)\rho_0^2(r),
\end{equation}
where $\rho_0^2\approx n_0$ for most positions inside the condensate. Here the pairs are
associated with radial wavenumber ($k,k^\prime$) and an angular momentum quantum number
$l=l_0+\Delta l$ and $l^\prime=l_0-\Delta l$. The total angular momentum is $2l_0\hbar$,
conserving the angular momentum of the two original condensate atoms.

\begin{figure}
\includegraphics[width=.5\textwidth]
{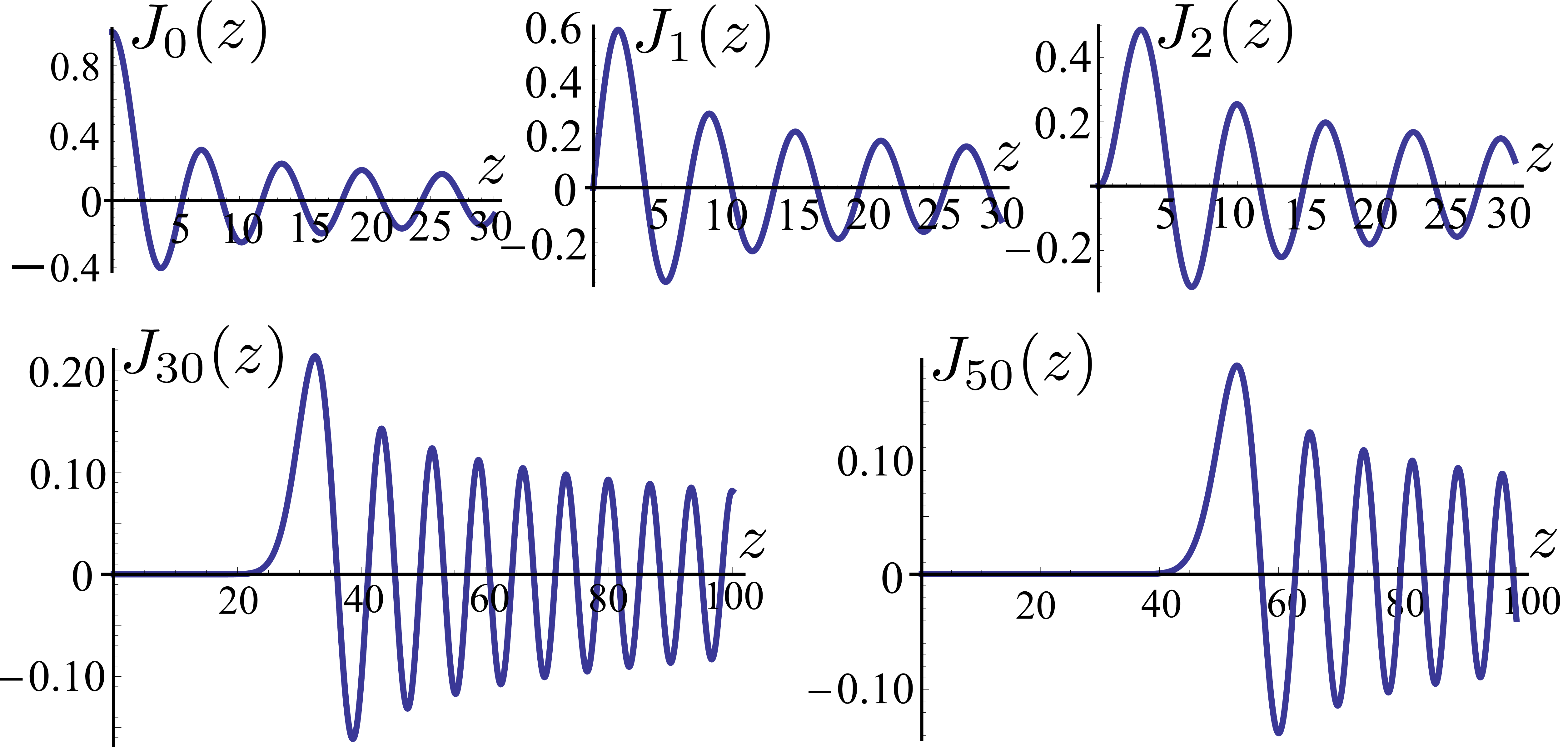}
\caption{Bessel functions values $J_l(z)$ as a function of the argument $z$ at orders $l=$0, 1, 2, 30, and 50.}
\label{fig:Bessel}
\end{figure}

To evaluate this integral, let us look at the properties of Bessel functions. For convenience, from now we choose $l_0\ge 0,\,\Delta l\ge 0$. A Bessel function $J_l(z)$ grows as $z^l$ at small $z$ and asymptotically behaves as $\cos(z-l\pi/2-\pi/4)/\sqrt{z}$ at large $z$. The transition happens at $z=g(l)$ where $J_l$ peaks. Empirically, it can be seen from the numerical plot of the Bessel functions that $g(l)$ increases monotonically with $l$ and is approximately $\propto l$ at large $l$ (see Fig. \ref{fig:Bessel}).
In our case, due to the high driving frequency, $l,l^\prime\gg1$. We estimate that the integral \eqref{eq:vortexF} peaks around $kR\sim l$ and $k^\prime R\sim l^\prime$, or
\begin{equation}\label{eq:dominantk}
k\sim |l_0+\Delta l|/R,\quad k^\prime\sim
|l_0-\Delta l|/R.
\end{equation}
This is consistent with the classical limit which is approached since $\Delta l\gg 1$. For the excited particle with  angular momentum $l\hbar$ at characteristic distance $\sim R$ from the trap center, the momentum is $\sim |l|\hbar /R$.
The energy conservation constraint, which can be derived from the resonance condition in the rotating wave approximation, imposes the constraint
\begin{equation} \label{eq:energy}
\frac{\hbar^2k^2}{2m}+
\frac{\hbar^2k^{\prime2}}{2m}=\simeq \hbar \omega.
\end{equation}
Because $\Delta l\gg l_0$, the parametrically amplified pairs then satisfy $k=k_f+\Delta k,\, k^\prime=k_F-\Delta k$ where $\Delta k\sim l_0/R$. This gives the two-ring structure in $k$-space distribution, see Fig. \ref{fig:vortexk}(a). And the scaling of $\Delta k$ with $l_0$ is numerically observed (see Fig. \ref{fig:vortexk}(b)). The outer ring $k$ has a phase winding in the same direction as $l_0$ while the inner ring $k^\prime $ has the opposite. Their sum gives 2$l_0$. All of these are verified as shown by Fig. \ref{fig:vortexk}(c).

As mentioned earlier, this pair generation explains the spiral jets arising from interference. In the far field, a Bessel function is asymptotically a cosine function and can be approximated by a mode propagating outward radially (the inward one can be ignored). The two $(k,\,k^\prime)$ modes then have a relative phase $2\Delta kr+2\Delta l\phi+\theta_i$ where $\theta_i$ is some initial relative phase. The constructive interference (of maximum density) occurs at relative phase equal to 2$\pi$ multiples and the trajectories then follow
$dr/d\phi=-\Delta l/\Delta k\sim -k_fR^2/l_0$, which is a spiral winding in the opposite direction of $l_0$.

\begin{figure}[h]
\includegraphics[width=.5\textwidth]
{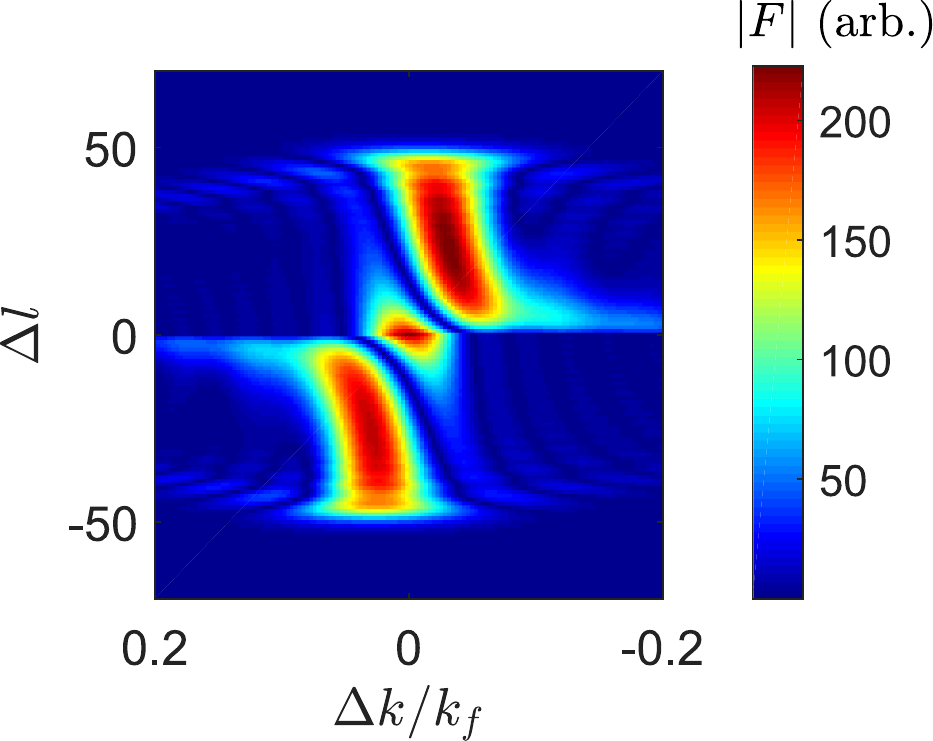}
\caption{$|F|$ at $l_0=1$. The absolute value of $F(l,l^\prime;k,k^\prime)$ peaks in several regions,  but only the one satisfying Eq. \eqref{eq:dominantk} has substantial area. Therefore, these modes dominate the particle occupation
states of the emitted jets. }
\label{fig:besselint}
\end{figure}

\begin{figure}[h]
\includegraphics[width=.5\textwidth]
{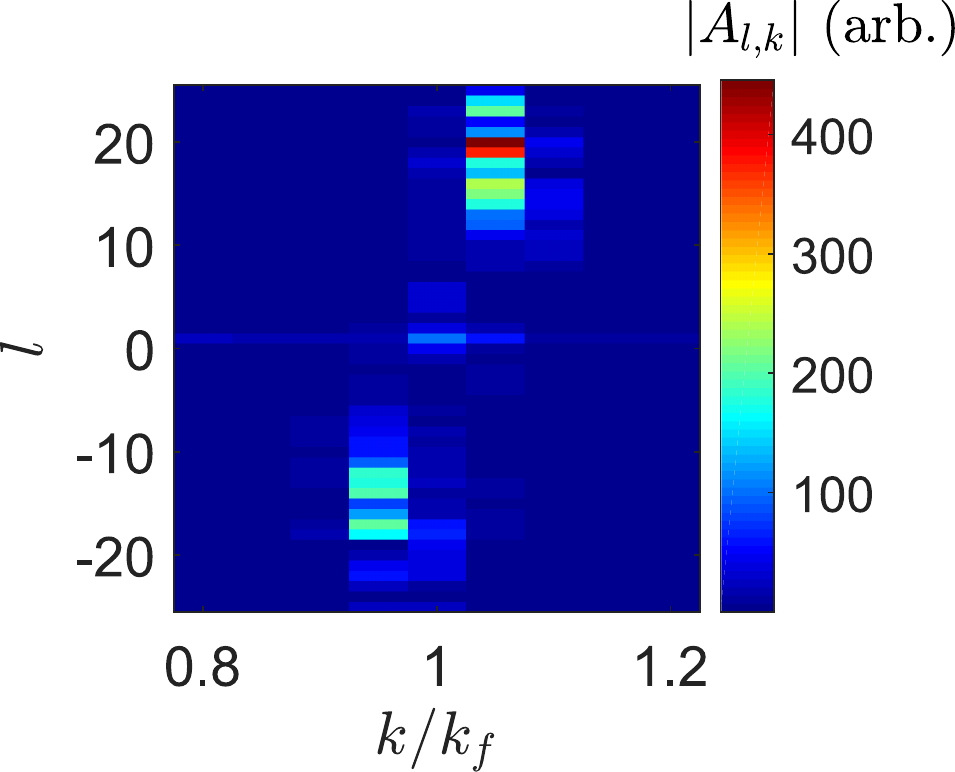}
\caption{Expansion in $(l,k)$ basis for wavefunction at $l_0=1$. The absolute value of the mode amplitude $A_{l,k}$ peaks at large $l$ and $k\neq k_f$, consistent with our analysis.}
\label{fig:expand}
\end{figure}

More quantitatively, we can directly evaluate the integral $F$ numerically similar to the soliton case. The results are presented in Fig. \ref{fig:besselint}, and are consistent with our simple arguments.
In particular, when we expand the wavefunction generated from our simulations in terms of the basis $(l,k)$, we see the particle occupation peaks near the resonance values $(k_f\pm\Delta k,\,l_0\pm\Delta l)$ consistent with our analysis, see Fig. \ref{fig:expand}.

One issue worth pointing out is that this pair generation mechanism still conserves the $0-\pi$ symmetry observed in the uniform condensate. Since $kR\lesssim l,\,k^\prime R\lesssim l^\prime$, we have $J_{l}(kr)\sim (kr)^{l}$ and $J_{l^\prime}\sim (-1)^{l^\prime}(k^\prime r)^{-l^\prime}$ where $l^\prime <0$. Therefore, the corresponding $F$ has a sign $(-1)^{l^\prime}$ which couples the amplitudes of these two modes.
\begin{equation}\label{eq:eom_V}
\begin{aligned}
i&\hbar \frac{\partial a_{l,k}(t)}{\partial t}-(\frac{\hbar^2k^2}{2m}+\mu)a_{l,k}(t)\\
\approx& -\frac{U_1e^{-i\omega t}}{2i}\delta k a_{l^\prime, k^\prime}^\dagger(t)F.\\
\end{aligned}
\end{equation}
This leads to the growth of the mode amplitude after the stimulation as
\begin{equation}\label{eq:coeff}
\begin{aligned}
A_{l,k}(t)=A_{l,k}(0)e^{\lambda t},\\
A_{l^\prime,k^\prime}(t)=(-1)^{l^\prime}A_{l, k}^*(t),
\end{aligned}
\end{equation}
where the growth exponent $\lambda\sim U_1\delta k |F |$ and the dynamical phase $\propto E_kt$ is not contained in this expression. This coefficient coupling results in the symmetry in the momentum space. The inner product between $\varphi_{\bf{q}}$ and $\varphi_{l,k}$ is
\begin{equation}
\begin{aligned}
\int d\vec{r}e^{-i \vec{q}\cdot \vec{r}}\sqrt{\frac{k}{2\pi}}e^{il\phi}J_l(kr)
=\sqrt{\frac{2\pi}{k}}e^{il(\phi_{\bf{q}}-\pi/2)}\delta(q-k).
\end{aligned}
\end{equation}
When there is a series of $l$-states centered around $l_r\approx k_fR$ which obey $A_{l,k}\approx A_{l_r,k}e^{-i \delta l (\phi_0-\pi/2)}$ where $\delta l =l-l_r$, we then find the amplitude for the plane wave mode $\bf{q}$ as
\begin{equation}
\begin{aligned}\label{eq:autopeak}
A_{\bf{q}}=&\sum_l  A_{l, k}\sqrt{\frac{2\pi}{k}}e^{il(\phi_{\bf{q}}-\pi/2)}\delta(q-k)\\
\approx&\sqrt{\frac{2\pi}{k}}A_{l_r, k}e^{il_r(\phi_{\bf{q}}-\pi/2)}\delta (q-k)\int d\delta l e^{i \delta l (\phi_{\bf{q}}-\phi_0)}\\
\approx&\sqrt{\frac{2\pi}{k}}2\pi A_{l_r, k}e^{il_r(\phi_{\bf{q}}-\pi/2)}\delta (q-k)\delta(\phi_{\bf{q}}-\phi_0).
\end{aligned}
\end{equation}
Meanwhile,we have the coupled $l^\prime$-states centered at $l_r^\prime=2l_0-l_r$ giving rise to
\begin{equation}
\begin{aligned}\label{eq:pipeak}
A_{\bf{q}}=&\sum_{l^\prime} A_{l^\prime, k^\prime}\sqrt{\frac{2\pi}{k^\prime}}e^{il^\prime(\phi_{\bf{q}}-\pi/2)}\delta(q-k^\prime)\\
=&\sqrt{\frac{2\pi}{k^\prime}}\delta(q-k^\prime)\sum_{l^\prime}(-1)^{l^\prime}A_{l=2l_0-l^\prime, k}^*e^{il^\prime(\phi_{\bf{q}}-\pi/2)}\\
\approx& \sqrt{\frac{2\pi}{k^\prime}} A_{l_r, k}^*e^{i l_r^\prime(\phi_{\bf{q}}+\pi/2)}\delta(q-k^\prime)\int d\delta l^\prime e^{i\delta l^\prime(\phi_{\bf{q}}-\phi_0+\pi)}\\
\approx&\sqrt{\frac{2\pi}{k^\prime}}2\pi A_{l_r, k}^*e^{i l_r^\prime(\phi_{\bf{q}}+\pi/2)}\delta(q-k^\prime)\delta(\phi_{\bf{q}}-\phi_0+\pi).\\
\end{aligned}
\end{equation}
This shows that when converting to the plane wave basis, there is always a coupled counter-propagating pair but now with different wavenumber $k,\,k^\prime$. This is verified by our numerical results (see Fig. \ref{fig:vortexk}(d)) where there is a perfect $0-\pi$ inversion symmetry between the $k_f\pm\Delta k$ ring.

\subsection{6.2. More quantitative description of near and far fields.}

We know that the uniform condensate can be regarded as a $l_0=0$ special example of the vortex case. Let us now deploy the angular basis to discuss several properties of the usual Bose fireworks. The generated pair is now $(l,k),\,(-l,k^\prime)$ with $l$ and $k$ peaked near $k_fR$ and $k_f$ and $k^\prime\approx 2k_f-k$. The wavefunction at an observation point $(r,\phi)$ is then
\begin{equation}
\begin{aligned}\label{eq:wavefn}
\psi(r,\phi)=&\sum_{l\ge 0}\int_{k\sim k_f} dk\left[A_{l,k}e^{-iE_k t}\sqrt{\frac{k}{2\pi}}J_l(kr)e^{il\phi}\right.\\
&\left.+A_{-l,k^\prime}e^{-iE_{k^\prime}t}\sqrt{\frac{k^\prime}{2\pi}}J_{-l}(k^\prime r)e^{-il\phi}\right]\\
\end{aligned}
\end{equation}
At large $r$ where $kr\gg l$ or $r\gg R$, the system is in the Fresnel regime, and the Bessel function has the asymptotic form \cite{Besselasy}
\begin{equation}
\begin{aligned}\label{eq:Besselasy}
J_l(kr)=&\sqrt{\frac{2}{\pi kr}}\left[\cos\left(kr-\frac{l\pi}{2}-\frac{\pi}{4}\right)P(l,kr)\right.\\
&\left.-\sin\left(kr-\frac{l\pi}{2}-\frac{\pi}{4}\right)Q(l,kr)\right]\\
P(l,kr)=&1-\frac{(4l^2-1)(4l^2-9)}{2!(8kr)^2}+\ldots,\\
 Q(l,kr)=&\frac{4l^2-1}{8kr}-\ldots.\\
\end{aligned}
\end{equation}
The remainder after $M$ terms in the expansion of $P(l,kr)$ doesn't exceed the $(M+1)$th term in absolute value and is of the same sign, provided that $M>l/2-1/4$. The same is true for $Q(l,kr)$ provided that $M>l/2-3/4.$ When the system is not in the Fraunhofer regime yet: $l\ll kr\ll l^2$, one can ignore the remainder after the $M$th term as the $M+1$th term is much smaller than unity here. For the $N$th term in $P(l,kr)$, where $1\le N\le M$, for most of the time, it can approximately be taken as $(-1)^N (4l^2)^{2N}/(2N)!(8kr)^{2N}$ as $l\gtrsim 2N$. So we have
$$P(l,kr)\sim \cos\left(\frac{4l^2}{8kr}\right),\quad Q(l,kr)\sim \sin\left(\frac{4l^2}{8kr}\right),$$
for which the higher terms after $M$th order is also negligible as $kr\gg l$.
Therefore, we arrive at the approximate expression of Bessel function as
\begin{equation}\label{eq:FresBessel}
J_l(kr)=\sqrt{\frac{2}{\pi kr}}\cos\left(kr-\frac{l\pi}{2}+\frac{l^2}{2kr}-\frac{\pi}{4}\right).
\end{equation}

As in the previous subsection, we can approximately take $A_{l,k}\simeq A_{l_r,k_f}e^{-i(\phi_0-\pi/2)\delta l-i r_0 \delta k}$ where $|r_0|<1/|\delta k|\simeq R$. The wavefunction is then
\begin{equation}
\begin{aligned}
\psi(r,\phi)\simeq&\sum_{l\ge 0}\int_{k\sim k_f} \frac{dk}{\pi\sqrt{ r}}\left[A_{l,k}e^{-iE_kt}\cos\left(kr-\frac{l\pi}{2}+\frac{l^2}{2kr}-\frac{\pi}{4}\right)e^{il\phi}\right.\\
&\left.+A_{l,k}^*e^{-iE_{k^\prime}t}\cos\left(k^\prime r-\frac{l\pi}{2}+\frac{l^2}{2kr}-\frac{\pi}{4}\right)e^{-il\phi}\right].\\
\approx &\frac{e^{-iE_{f}t}}{2\pi\sqrt{ r}}\sum_{l\ge 0}\int_{k\sim k_f} dk \left\{ A_{l_r,k_f}
e^{-i(v_ft +r_0) \delta k-i(\phi_0-\frac{\pi}{2})\delta l}\right.\\
&\left. \left[e^{i(k_fr-\frac{l_r\pi}{2}+\frac{l_r^2}{2k_fr}-\frac{\pi}{4}+l_r\phi)}e^{ir\delta k +i(\phi-\frac{\pi}{2}+\frac{l_r}{k_fr})\delta l}\right.\right.\\
&\left.+e^{-i(k_fr-\frac{l_r\pi}{2}+\frac{l_r^2}{2k_fr}-\frac{\pi}{4}-l_r\phi)}e^{-ir\delta k +i(\phi+\frac{\pi}{2}-\frac{l_r}{k_fr})\delta l)}\right]\\
&+ A_{l_r,k_f}^*e^{i(v_ft+r_0 )\delta k+i(\phi_0-\pi/2)\delta l} \\
&\left[e^{i(k_fr-\frac{l_r\pi}{2}+\frac{l_r^2}{2k_fr}-\frac{\pi}{4}-l_r\phi)}e^{-ir\delta k-i(\phi+\frac{\pi}{2}-\frac{l_r}{k_fr})\delta l}\right.\\
&\left.\left.+e^{-i(k_fr-\frac{l_r\pi}{2}+\frac{l_r^2}{2k_fr}-\frac{\pi}{4}+l_r\phi)}e^{ir\delta k-i(\phi-\frac{\pi}{2}+\frac{l_r}{k_fr})\delta l}\right]\right\}\\
\approx& \frac{2\pi}{\sqrt{ r}}e^{-i(E_{f}t-k_fr-\frac{l_r^2}{2k_fr}+\frac{\pi}{4})}\delta(r-v_ft-r_0)\\
&\left[A_{l_r,k_f}e^{il_r(\phi-\frac{\pi}{2})}\delta \left(\phi-\phi_0+\frac{l_r}{kr}\right)+\right.\\
&A_{l_r,k_f}^*e^{-il_r(\phi+\frac{\pi}{2})}\delta\left(\phi-\phi_0+\pi-\frac{\l_r}{kr}\right).\\
\end{aligned}
\end{equation}
The radially inward propagation mode is neglected since $v_ft+r+r_0\gg0$ at $r\gg R$.
The density is no longer symmetric at $\pi$ relative angle, but shifts by a random angle $\sim l_r/kr\sim R/r$ which is roughly the angular span of the condensate relative to the measurement point. This is consistent with the intuitive picture in Ref. \cite{DensityWave_supp} but presented more quantitatively here. This can be inferred from the picture that the generated pair is always opposite but their connecting line can be away from the trap center. We infer that the relative angle with respect to the trap center (the origin) is smaller than $\pi$ by $\sim R/r$. This is the near-field asymmetry.

The derivations here ignore second order corrections to the phase $\Delta l^2/k r\sim l^2/kr>1$. To be more accurate, one should write the wavefunction as a summation over several wave packets. But more rigorous arguments will give the same conclusions presented above.

When the system is in the Fraunhofer regime ($kr\gg l^2$ or $R/r\ll 1/l$), one can ignore all the higher orders in Eq. \eqref{eq:Besselasy} and we find
\begin{equation}
\begin{aligned}
\psi(r,\phi)\simeq&\sum_{l\ge 0}\int_{k\sim k_f} \frac{dk}{\pi\sqrt{r}}\left[A_{l,k}e^{-iE_kt}\cos\left(kr-\frac{l\pi}{2}-\frac{\pi}{4}\right)e^{il\phi}\right.\\
&\left.+A_{l,k}^*e^{-iE_{k^\prime}t}\cos\left(k^\prime r-\frac{l\pi}{2}-\frac{\pi}{4}\right)e^{-il\phi}\right].\\
\approx &\frac{2\pi}{\sqrt{ r}}e^{-i(E_{f}t-k_fr+\frac{\pi}{4})}\delta(r-v_ft-r_0)\\
&\left[A_{l_r,k_f}e^{il_r(\phi_0-\frac{\pi}{2})}\delta (\phi-\phi_0)+\right.\\
&\left.A_{l_r,k_f}^*e^{-il_r(\phi_0-\frac{\pi}{2})}\delta(\phi-\phi_0+\pi)\right]\\
\end{aligned}\label{eq:fraunhofer}
\end{equation}
which recovers the $\pi$ peak symmetry in the so-called far field, and has a correlation width $\sim 1/\Delta \delta l\sim1/k_fR$.

One should note here that due to the finite condensate size, the system has only cylindrical symmetry with translational symmetry lost. Therefore, the angular width is constant throughout the jet time of flight while the angular linear width is expanding with time. At first sight, this seems contradictory with that the plane wave mode has angular correlation width $1/\Delta\delta l\sim 1/k_fR$ (Eqs. \eqref{eq:autopeak} and \eqref{eq:pipeak}). However, one should pay attention to the expansion of $E_qt\approx E_ft+v_ft\Delta |{\bf{q}}|$ where $\Delta |{\bf{q}}|=\sqrt{(q+\Delta q)^2+(q\Delta \phi_{\bf{q}})^2}-q\sim \Delta q+q\Delta \phi_{\bf{q}}^2/2$. In the far field when $v_ft=r\gg lR$, since each plane wave mode has an angular width as $\Delta \phi_{\bf{q}}\sim 1/k_fR$ and $q\approx k_f$, we find that the second order contribution to the dynamical phase has exceeded order unity in the far field $k_fr\gg (k_fR)^2$. This leads to the linear expansion of the wave packet or the jet in the angular direction and is unrelated to the nonlinear energy dispersion.

\section{7. Geometric analysis of the multiple-slit interference}
Lastly, we present the geometric argument behind the intuitive multiple-slit interference picture.
\begin{figure}
\includegraphics[width=.2\textwidth]
{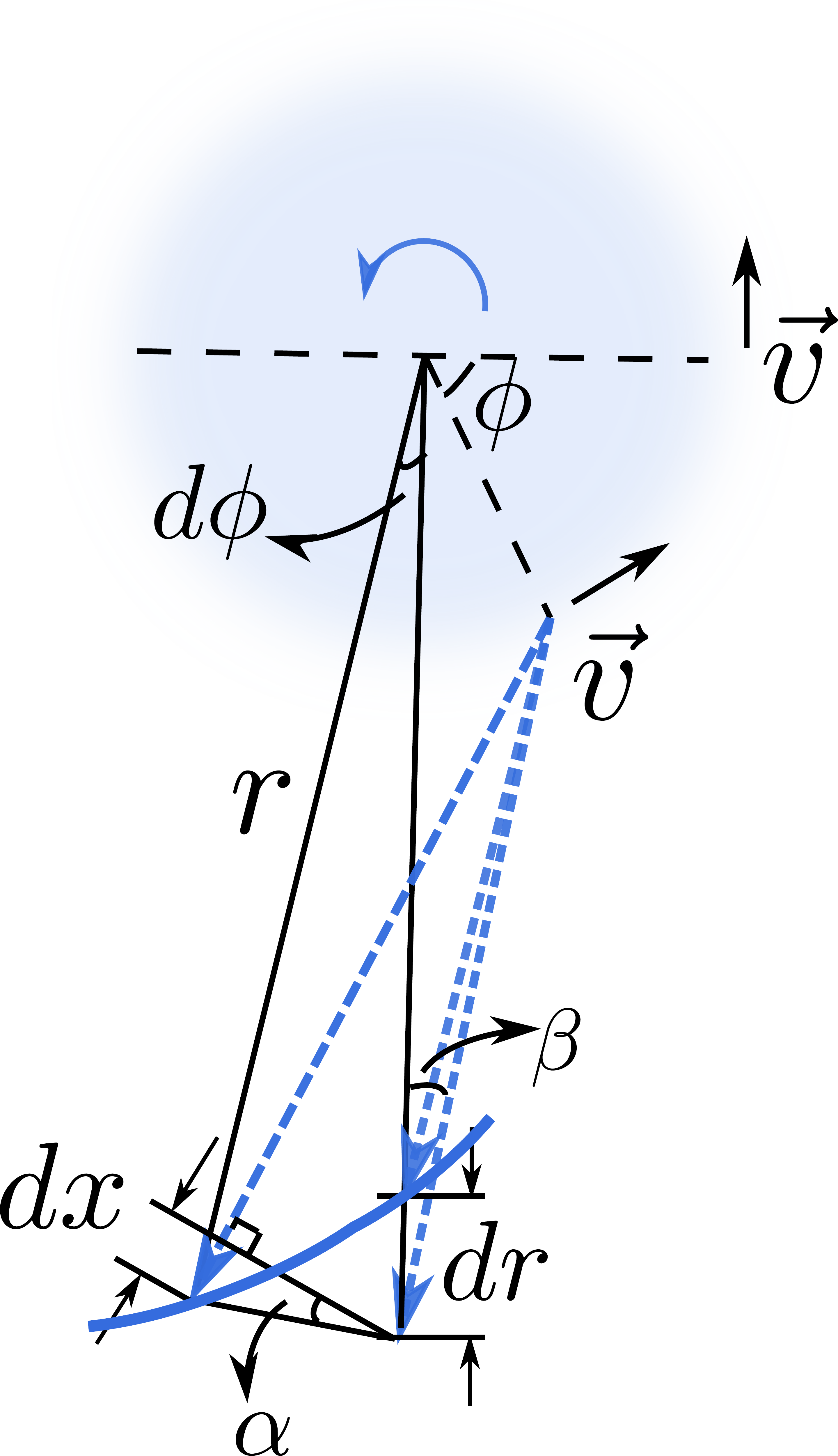}
\caption{Schematic of spiral trajectories resulting from interference between emission from different parts of a vortex-embedded BEC.}
\label{fig:schematic1}
\end{figure}
As mentioned in the main text, each point in the trap functions as an individual source emitting different modes the wavenumbers of which are $k_f+ (mv/\hbar)\cos(\phi_r)$ dependent on the relative angle $\phi_r$ between final and initial velocities.
For an observation point at distance $r\gg R$ from the trap center, there are jets emitted from different ``sources" overlapping at this point, see Fig. \ref{fig:schematic1}. If the jet comes from a point at angular position $\phi$ ($\phi$ is measured with respect to the axis perpendicular to the line connecting the trap center and the measurement point), the jet wavenumber is then $k_f-(mv/\hbar) \cos\phi$ for each source. Therefore, when the measurement point shifts by a radial distance $dr$, the optical paths for the jets from different sources would all increase as $\sim dr$ but have different phase accumulations due to different $k$ values. To keep the relative phases between different modes unchanged, the observation point needs to shift an angle of $d \phi$ so that the modes with
larger $k$ values would have shorter optical paths (see Fig. \ref{fig:schematic1}). In this way, the optical path changes by $dx$ which can be easily derived from geometric analysis:
$$dx\approx r d\phi\sin \alpha, $$ where $$\alpha\approx \beta,\quad r\sin\beta=R\cos\phi.$$
Therefore, the total phase accumulation for each mode is approximately $$\left(k_f-\frac{mv}{\hbar}\cos\phi\right)dr+k_fR\cos\phi d\phi,$$ which is a constant for all modes only when
$$\frac{mv}{\hbar}dr=k_fRd\phi. $$ Since the interference fringe is along fixed relative phases, we then obtain the trajectory as $$\frac{dr}{d\phi}\approx\frac{k_fR^2}{l_0},$$ where $mv/\hbar\sim l_0/R$. This spiral winds in the opposite direction to that of the original vortex, as seen from Fig. \ref{fig:schematic1}.

\end{document}